\documentclass[preprint,aps,superscriptaddress,nofootinbib,eqsecnum]{revtex4}
\usepackage{psfig}

\let\ga=\gamma

\let\De=\Delta

\let\del=\nabla
\let\si=\sigma

\let\Om=\Omega

\def\to{\rightarrow}
\let\p=\partial
\let\<=\langle
\let\>=\rangle

\let\txt=\textstyle

\def\beq{\begin{equation}}
\def\eeq{\end{equation}}
\def\ba{\begin{array}}
\def\bea{\begin{eqnarray}}
\def\ea{\end{array}}
\def\eea{\end{eqnarray}}

\def\slash{\!\!\!\!/\,}

\def\comment#1{ \hbox{[{\it Comment suppressed here.}\/]} }
\def\hide#1{}

\def\IR{\relax{\rm I\kern-.18em R}}
\def\IN{\relax{\rm I\kern-.18em N}}
\def\IB{\relax{\rm I\kern-.18em B}}
\def\IE{\relax{\rm I\kern-.18em E}}
\def\ZZ{\relax{\sf Z\kern-.4em Z}}     % use as $\ZZ$ or $\ZZ_p$

\def\ontopss#1#2#3#4{\raise#4ex \hbox{#1}\mkern-#3mu {#2}}

\newcommand{\skipover}[1]{}
\newcommand{\nn}{\nonumber \\}

\def\half {{\txt {1\over 2}}}

\def\pbar{{\bar p}}
\def\phat{\hat{p}}

\def\kbar{{\bar k}}
\def\khat{\hat{k}}
\def\khatbf{\hat{\bf k}}
\def\psib{{\bar\psi}}
\def\Omb{{\bar\Om}}

\def\={\!=\!}
\def\+{\,+\,}
\def\-{\,-\,}

\newcommand{\xiinf}{\xi \to \infty}
\newcommand{\static}{m_Q \to \infty}

\newcommand{\atzero}{a_t \to 0}
\newcommand{\sA}{A}
\newcommand{\sB}{B}
\newcommand{\sC}{C}

\newcommand{\fB}{{F_B}}
\newcommand{\fC}{F_C}
\newcommand{\fBbarmu}{\bar{F}_{B\mu}}
\newcommand{\fCbar}{\bar{F}_C}
\newcommand{\Lbar}{\bar{L}}
\newcommand{\Kbar}{\bar{K}}
\newcommand{\Xbar}{\bar{X}}
\newcommand{\Ybar}{\bar{Y}}
\newcommand{\aX}{a_\mu X_\mu}
\newcommand{\aY}{a_\mu Y_\mu}
\newcommand{\aZ}{a_\mu Z_\mu}
\newcommand{\aW}{a_\mu W_\mu}
\renewcommand{\ap}{a_\mu p_\mu}

\newcommand{\intk}{\int_{\bf k}}
\newcommand{\intinf}{\int^{\infty}_{-\infty}}
\newcommand{\gprop}{\frac{1}{\xi^2 \khat_0^2 + \khatbf^2}}

\newcommand{\pbf}{{\bf p}}
\newcommand{\kbf}{{\bf k}}
\newcommand{\zbf}{{\bf 0}}
\newcommand{\pzero}{\pbf = \zbf}
\newcommand{\on}{i M_1,\zbf}

\newcommand{\inatzero}{~\stackrel{\atzero}{\longrightarrow}~}
\newcommand{\instatic}{~\stackrel{\static}{\longrightarrow}~}
\newcommand{\Ogf}{O(g^4)}
\newcommand{\sub}{{\rm sub}}
\newcommand{\nosub}{{\rm nosub}}
\newcommand{\reg}{{\rm reg}}
\newcommand{\tad}{{\rm tad}}
\newcommand{\TI}{{\rm T.I.}}
\newcommand{\ig}{\frac{1}{a+bx^2}}
\newcommand{\iqn}{ie-x}
\newcommand{\iqd}{g^2(ie-x)^2+f^2}
\begin{document}

\preprint{KEK-CP-137}
\preprint{FERMILAB-Pub-03/030-T}

\title{Anisotropic lattice actions for heavy quark}

\newcommand{\kek}{
  High Energy Accelerator Research Organization (KEK),
  Tsukuba, Ibaraki 305-0801, Japan}

\newcommand{\fnal}{
  Theoretical Physics Department, 
  Fermi National Accelerator Laboratory,
  P.O. Box 500, Batavia, Illinois 60510, USA}

\author{Shoji Hashimoto}
\affiliation{\kek}

\author{Masataka Okamoto}
\affiliation{\kek}
\affiliation{\fnal}

\date{\today}

\begin{abstract}
  The anisotropic lattice fermion actions are investigated
  with the one-loop perturbative calculations aiming at
  constructing a formulation for heavy quark with controlled
  systematic uncertainties.
  For the heavy-light systems at rest the anisotropic
  lattice with small temporal lattice spacing $a_t$
  suppresses the discretization error by a power of 
  $a_t m_Q$ for a heavy quark of mass $m_Q$.
  We discuss the issue of large
  discretization errors, which scales as $a_s m_Q$ with
  $a_s$ the spatial lattice spacing.
  By performing one-loop calculations of the speed-of-light
  renormalization for several possible lattice actions in
  the limit of $a_t\rightarrow 0$, we show that one can
  eliminate the large systematic error on the anisotropic lattice.
\end{abstract}
%\pacs{PACS number(s): 11.10.Gh, 11.15.Ha, 12.38.Bx, 12.38.Gc }

\maketitle

\section{Introduction}
\label{sec:intro}
In the heavy quark physics, the lattice simulation of
Quantum Chromodynamics (QCD) is an indispensable tool to
compute hadron masses and matrix elements nonperturbatively
without introducing model dependence.
One of the most important hadron matrix elements in the $B$
physics is the $B$ meson decay constant $f_B$, for which a
number of lattice calculations have been performed so far
and the systematic uncertainties are under control at the
level of 15\% accuracy \cite{Yamada:2002wh}.
In future, further precise calculation, say better than 5\%,
is necessary to constrain the Cabibbo-Kobayashi-Maskawa
(CKM) matrix elements more strictly and to search for the
signature of new physics.

One of the dominant uncertainties in the lattice simulation
of heavy quark is the systematic error associated with the
large heavy quark mass, since the lattice cutoff $1/a$
available with current computer power is not much larger
than the heavy quark mass $m_Q$.
A conventional approach to avoid this problem is to restrict
ourselves in the region where the systematic error is under
control ($m_Q\ll 1/a$) and to extrapolate to the $b$ quark
mass using the heavy quark scaling law predicted by the
heavy quark effective theory (HQET).
This is unsatisfactory in order to achieve the 5\% accuracy,
since the possible error scales as $(am_Q)^n$ ($n$ = 2 for
the $O(a)$-improved action) and thus grows very quickly
toward heavier quark masses.
The extrapolation to the $b$ quark mass could even amplify
the systematic uncertainty.

Another method is the HQET-based approach which includes 
lattice NRQCD \cite{Thacker:1990bm,Lepage:1992tx} and the
Fermilab method \cite{El-Khadra:1997mp}.
In this method one considers the lattice action for heavy
quark as an effective theory valid for large heavy quark
masses.
The advantage of the HQET-based approach is the absence of
the large systematic error which scales as $(am_Q)^n$.
The price one has to pay, on the other hand, is the
introduction of a number of terms in the action.
Their associated coefficients have to be determined by
matching the effective theory onto the continuum full
theory. 
The matching is usually carried out using perturbation
theory, which limits the accuracy of the lattice
calculation. 

Besides the HQET-based approach, a possible way to control 
heavy quark discretization effects
is to consider an anisotropic lattice, where the temporal
lattice spacing $a_t$ is much smaller than the spatial one
$a_s$ \cite{Alford:1996nx,Klassen:1999fh}. 
Since for a heavy meson (or a heavy baryon) at rest the
large energy scale of order $m_Q$ appears only in the
temporal component in the momentum space, one can expect
that the systematic error arises as $(a_t m_Q)^n$
and therefore suppressed as far as $a_t$ is small enough.
The computational cost is not prohibitive if one keeps
the spatial lattice spacing relatively large.
The problem of the matching of many operators in the
effective theory does not appear, as the theory is 
relativistic.

There is, however, a subtle issue discussed in
\cite{Harada:2001ei} that for a certain choice of the Wilson
term in the spatial direction the systematic error may arise
in the combination $a_s m_Q$ rather than the expected 
$a_t m_Q$ and the virtue of the anisotropic lattice is
spoiled.
With an alternative choice the error of order 
$(a_s m_Q)^n$ may be avoided but the unwanted doublers
become lighter and disturbs the simulation of physical
states.
The authors of \cite{Aoki:2001ra} even denied the advantage
of the anisotropic lattice used for heavy quark based on
their observation of $a_s m_Q$-like behavior through
radiative corrections. 
In this paper we discuss this issue further by considering a
larger set of $O(a)$-improved lattice fermion actions and
by performing one-loop calculations in the limit
$a_t\rightarrow 0$ where no $a_t m_Q$ error remains.

The appearance of large systematic errors scaling
as $a_s m_Q$ is naively unexpected for the following
reasons. 
In the $a_t\rightarrow 0$ limit the only source of the
discretization error is the spatial derivative in the
lattice action.
In momentum space, therefore, discretization errors
scale as $a_s p$ with $p$ a typical (spatial) momentum
scale in the system, which is of order
$\Lambda_{\mathrm{QCD}}$ for the heavy-light mesons or
baryons at rest, and the combination $a_s m_Q$ may not
appear as the momentum of order $m_Q$ flows only in the
temporal direction.
This intuitive picture should be correct even after 
radiative corrections, because the large momentum of order
$m_Q$ does not flow into the spatial direction in the
momentum space, and therefore the discretization error in
the spatial lattice derivative cannot accompany the heavy
quark mass $m_Q$.
It becomes clearer if one considers the limit 
$1/a_s \ll m_Q \ll 1/a_t$, because the spatial momentum
integral runs up to $\pi/a_s$ and thus cannot pick up the
larger scale $m_Q$.

Here, in order to understand the reason why the 
unexpected $a_s m_Q$-type error may appear in
\cite{Harada:2001ei,Aoki:2001ra},
let us consider the energy-momentum dispersion
relation at the tree level. 
We consider the $a_t\rightarrow 0$ limit, and the spatial
lattice spacing is also kept small enough such that we can
neglect the error of $O(a_s^2)$ and higher.
For the Wilson-type fermions the inverse quark propagator is
given as 
\begin{equation}
  m_Q + i\gamma_0 p_0 + i \sum_i\gamma_i p_i
  + \frac{r_s}{2\xi} a_s \sum_i p_i^2 + O(a_s^2),
\end{equation}
where $r_s$ denotes the coefficient in front of the spatial
Wilson term as defined in (\ref{D234}) in the next section.
The term including the anisotropy $\xi=a_s/a_t$ is
maintained even in the $a_t\rightarrow 0$ limit, because
that term could remain when $r_s$ scales as $\xi$.
To push up the spatial doubler mass in the cutoff scale,
$r_s/\xi$ must be kept finite.

The energy-momentum dispersion relation becomes
\begin{eqnarray}
  - p_0^2 
  & = &
  \left( m_Q + \frac{r_s}{2\xi} a_s \sum_i p_i^2 \right)^2
  + \sum_i p_i^2 + O(a_s^2)
  \nonumber\\
  & = &
  m_Q^2 + 
  \left(1 + \frac{r_s}{\xi} a_s m_Q\right) \sum_i p_i^2 
  + O(a_s^2),
\end{eqnarray}
and thus the error of order $a_s m_Q$ appears unless
$r_s/\xi$ vanishes, for which the doublers become light. 
Since the term $(r_s/\xi) a_s m_Q$ comes from the cross term
of the mass and the Wilson terms, the origin of the
combination $a_s m_Q$ is nothing to do with the large
momentum flow into the spatial direction.
At the tree level we may consider a set of lattice actions
in which there is no spatial Wilson term by introducing
higher derivative operators to decouple the unwanted
doublers. 
This class of actions does not have the problem of the 
$O(a_s m_Q)$ error at the tree level and may be used for
heavy quark even for $a_s m_Q > 1$.
There are higher order terms whose coefficient behaves like
$a_s m_Q$, but we neglect them as their contribution is
$O(a_s^2)$ or higher.

The problem is, then, whether the nice property of these
actions is maintained even with radiative corrections.
In this paper we perform one-loop perturbative calculation
for these lattice actions and investigate the mass
dependence of the rest mass $M_1=E(\zbf)$ and the kinetic
mass $M_2=({\partial^2E / \partial p_1^2})^{-1}_{\pzero}$,
where $E(\pbf)$ is the energy of heavy quark on-shell.
We examine the functional dependence of the speed of light 
renormalization parameter $\nu$, which is defined such that 
the relation $M_1=M_2$ is satisfied.
If the one-loop coefficient behaves as $(a_s m_Q)^n$,
the action suffers from the unwanted heavy quark mass
dependent error.
Because we are interested only in the $O((a_sm_Q)^n)$
errors, we carry out the one-loop calculation in the
$\atzero$ limit, where $O((a_tm_Q)^n)$ errors vanish. 
The fermion actions we consider are the anisotropic SW
(Sheikholeslami and Wohlert) action  
\cite{Sheikholeslami:1985ij,Harada:2001ei} 
and some special cases of the D234 action
\cite{Alford:1996nx}. 
We find the latter to be useful for applications to 
heavy quark systems.

This paper is organized as follows. 
In Section~\ref{sec:action}, we define the anisotropic
fermion actions we consider in this paper, and discuss their
tree-level properties. 
The static limit of those actions is considered in
Section~\ref{sec:limitorder}.
The one-loop calculation is then given in
Section~\ref{sec:oneloop}, whose results are presented in
Section~\ref{sec:results}.
Section~\ref{sec:concl} is devoted to our conclusions.
Some technical details are deferred to the Appendices.

\section{Anisotropic lattice fermion action}
\label{sec:action}
We start with the D234 quark action on the anisotropic 
lattice \cite{Alford:1996nx} given by
\beq
\label{eq:action}
S_{\rm D234} = a_t a_s^3 \sum_x \psib(x) Q \psi(x)
\eeq
\bea
\label{D234}
Q &=&  m_0 \+ \sum_\mu\,
      \nu_\mu \ga_\mu \del_\mu \, ( 1 - b_\mu a_\mu^2 \De_\mu ) \nn
&& \- {1\over 2} \, a_t  \,
   \Bigl( \sum_\mu r_\mu \Delta_\mu  \, \+ \, 
    \, \sum_{\mu < \nu} c_{\rm SW}^{\mu} \si_{\mu\nu} F_{\mu\nu} \Bigr)
            \+ \sum_\mu \nu_\mu d_\mu a_\mu^3 \De_\mu^2 
\eea
where $a_0=a_t$ and $a_i=a_s~(i=1,2,3)$ are the temporal and
spatial lattice spacings respectively, and 
\beq
(\nu_0,\nu_i) =(1,\nu)\, , \,\,\,\,\,\,  
(b_0,b_i) =(b_t,b_s)\, , \,\,\,\,\,\,   (r_0,r_i) =(r_t,r_s)\, ,
\eeq
\beq
(c_{\rm SW}^{0},c_{\rm SW}^{i}) = (c_{\rm SW}^{t},c_{\rm SW}^{s}) 
\, , \,\,\,\,\,\,  (d_0,d_i) =(d_t,d_s)\, . \,\,\, 
\eeq
Note that the lattice spacing in front of the Wilson and
clover terms is $a_t$, not $a_\mu$. 
This notation is similar to the one in \cite{Alford:1996nx},
but different from those in
\cite{Klassen:1999fh,Harada:2001ei}. 
The anisotropy parameter is defined by
\beq
\xi\equiv a_s/a_t .
\eeq
The lattice covariant derivatives 
$\del_\mu$, $\De_\mu$, $\del_\mu\De_\mu$ and $\De_\mu^2$
represent $D_\mu$, $D_\mu^2$, $D_\mu^3$ and $D_\mu^4$,
respectively, in the continuum theory, and their detailed
definitions are given in Appendix~\ref{sec:def}. 

In this paper we always set 
\beq
r_t=1\, , \ \ \ b_t = d_t =0\, .
\label{sD234}
\eeq
Thus, the operator $Q$ is nothing but the Wilson-Dirac
operator as far as the temporal derivatives are concerned. 
With this condition, the energy-momentum relation for the
fermion has a physical solution only, and the unphysical
temporal doublers do not appear \cite{Alford:1996nx}. 

Solving $Q(p)Q(p)^\dagger =0$ in the momentum space and then
setting $p_0=iE$, we obtain the energy-momentum relation for
the D234 action as 
\beq
4 \sinh^2({a_t E\over 2})
~=~ \frac{\nu^2 a_t^2 \sum_i \bar{p_i}^2 (1+b_s a_i^2 \hat{p_i}^2 )^2 
      + \mu({\bf p})^2}{1+\mu({\bf p})} \, , 
\label{disp_sD234}
\eeq
where
\bea
\mu({\bf p}) &=& a_t m_0 + {1\over 2} r_s a_t^2 \sum_i \phat_i^2 +
                          \nu d_s a_t \sum_i  a_i^3 \phat_i^4 ~,
\label{mu}
\eea
and $\pbar_\mu$ and $\phat_\mu$ are defined in
Appendix~\ref{sec:def}. 

From (\ref{disp_sD234}), we obtain the tree-level rest mass
$M_1=E(\zbf)$ 
and kinetic mass $M_2=({\partial^2E / \partial p_1^2})^{-1}_{\pzero}$ as 
\begin{eqnarray} 
a_tM_1 &=& \log (1+a_tm_0)\, ,\label{polemass}\\
\frac{1}{a_tM_2} &=& \frac{2\nu^2}{a_t m_0(2+a_tm_0)}
+ \frac{r_s}{1+a_tm_0}\, . \label{kineticmass}
\end{eqnarray}
On the anisotropic lattice with $a_tm_0 \ll 1$,
the tree-level mass ratio $M_1/M_2$ can be expanded in terms
of $a_tm_0$ as 
\beq
M_1/M_2 = 1 + (r_s-1) \, a_tm_0 + O((a_tm_0)^2) \, ,
\label{m1m2}
\eeq
where we set $\nu=1$.
The deviation of $M_1/M_2$ from unity is a lattice
discretization error arising from the fermion mass. 
Unless $r_s \propto \xi$, 
such an error is a function of $a_tm_0$ alone, 
which is small on the anisotropic lattice 
with $a_tm_0 \ll 1$ \cite{Harada:2001ei}. 
When $r_s \propto \xi$, a discretization error of order 
$a_sm_0=\xi a_tm_0$ arises, which is still large on the
anisotropic lattice. 

From (\ref{disp_sD234}), we can also calculate the
``spatial-doubler'' mass $E^d$,
\textit{i.e.} 
the energy at the edge of the Brillouin zone ($p_i=\pi/a_s$):
\bea
a_t E^d &=& \log \biggl[1+ a_t m_0 + 
\frac{2 r_s}{\xi^2}n_d + \frac{16\nu d_s}{\xi}n_d\biggr]\\
&\equiv& \log [1+ a_t m_0^d ] \, ,
\eea
where $n_d$ (= 1, 2, 3) is the number of spatial direction
with $p_i=\pi/a_s$, and the bare spatial-doubler mass
$m_0^d$ is given by
\beq
a_s m_0^d = a_s m_0 + \frac{2 r_s}{\xi}n_d + 16\nu d_s n_d \, 
\label{asm0d}
\eeq
in units of the spatial lattice spacing $a_s$.
We note that one has to take $r_s \propto \xi$ or $d_s > 0$
in order to decouple the spatial-doubler with the energy
$E^d$ from the physical state for large values of $\xi$.

It is interesting to consider the energy-momentum relation
in the Hamiltonian limit $\atzero$ ($\xiinf$), where
$a_tm_0$ errors vanish. 
In this limit the left-hand side of (\ref{disp_sD234}) is
replaced by $a_t^2E^2$, and the energy-momentum relation is
simplified to 
\bea
E^2(\atzero)
&\!=\!& \nu^2 \sum_i \bar{p_i}^2 (1+b_s a_i^2 \hat{p_i}^2 )^2  
      + \frac{\mu({\bf p})^2}{a_t^2} \\
&\!=\!& m_0^2 + \nu^2 {\bf p}^2 + 
\biggl(
-\frac{\nu^2}{3}+2 \nu^2 b_s + 2 a_s m_0  \nu d_s \biggr)a_s^2 \sum_i p_i^4
+ O(a^4 p^6) \, .
\label{disp_sD234_a0=0}
\eea
From a small $\pbf$ expansion we obtain a non-relativistic
expression of $E(\atzero)$:
\beq
E(\atzero) = m_0 + \frac{\nu^2}{2m_0} {\bf p}^2 
- \frac{\nu^4}{8m_0^3} ({\bf p}^2)^2
+ \frac{1}{2m_0} \biggl(
-\frac{\nu^2}{3}+2 \nu^2 b_s + 2 a_s m_0  \nu d_s \biggr)a_s^2 \sum_i p_i^4
+ O(a^5 p^6) \, .
\label{ENR}
\eeq
Taking the static limit $m_0 \to \infty$ subsequently, one
obtains 
\beq
E(\atzero) ~\stackrel{m_0 \to \infty}{\longrightarrow}~ 
m_0 + \nu d_s a_s^3 \sum_i p_i^4 + O(a^5 p^6) \, .
\label{Estat}
\eeq
Note that the $\sum_i p_i^4$ term survives even in the static limit.
This will be discussed later. 

We are now ready to define our anisotropic actions more
explicitly. 
We study two actions: one is the SW action
\cite{Sheikholeslami:1985ij,Harada:2001ei}, and the other is 
a variant of the D234 action \cite{Alford:1996nx}, which we
call the sD34 action. 
We give these actions and discuss their tree-level
properties in the following.

\subsection{SW action}
\begin{figure}[t]
\centerline{
\leavevmode\psfig{file=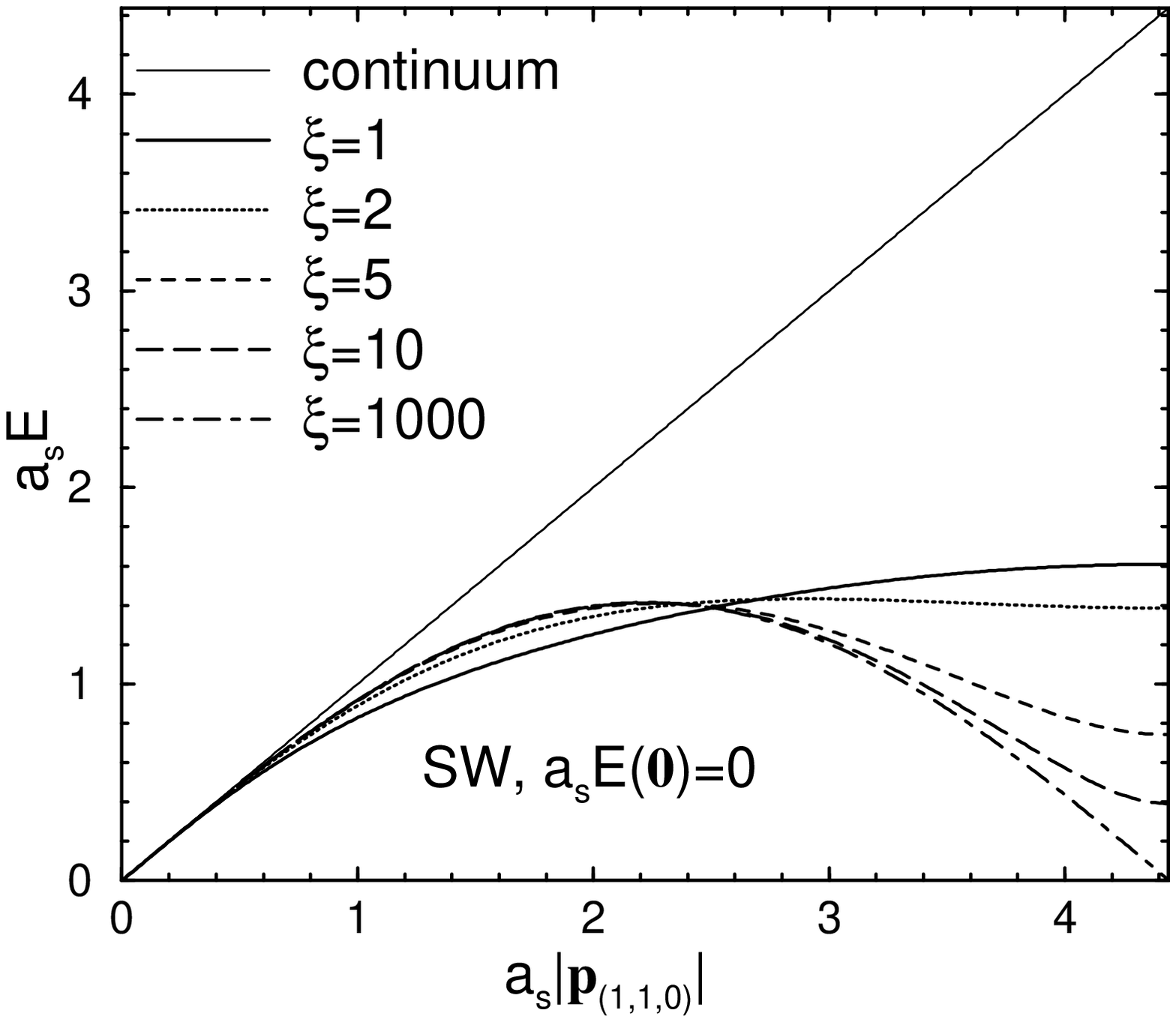,width=8.cm}
\leavevmode\psfig{file=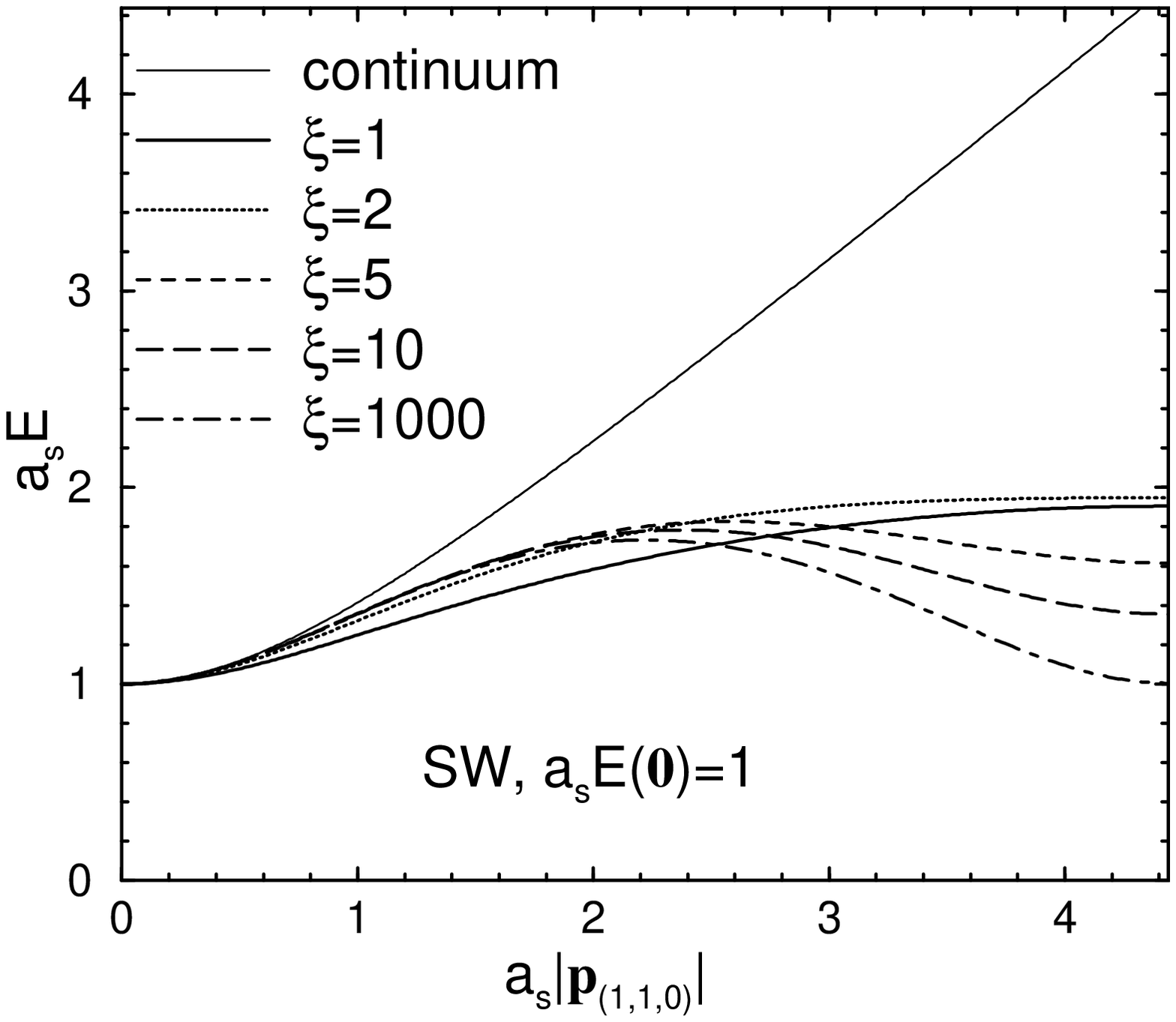,width=8.cm}}
\caption{
  Energy-momentum relation at different values of $\xi$ for
  the SW action.
  The left panel shows the case of $a_sE(\zbf)=0$, while the
  right shows $a_sE(\zbf)=1$.
  The spatial momentum $\pbf$ is along the $(1,1,0)$ direction. 
  For comparison we also plot the energy-momentum relation
  in the continuum.
}
\label{fig:disp_SW}
\end{figure}

The Sheikholeslami-Wohlert (SW) action
\cite{Sheikholeslami:1985ij,Harada:2001ei} is defined by 
\beq
\nu = r_s = c_{\rm SW}^{\mu} = 1  \, , \ \ b_s=d_s=0\, .
\label{eq:anisoW}
\eeq
An $O(a_t)$ error arising from the Wilson terms is removed
by $c_{\rm SW}^{\mu} = 1$.
Since the Wilson terms are $O(a_t)$, this action goes over
to the ``naive'' quark action in the $\atzero$ ($\xiinf$)
limit.
The energy splitting between the physical state and
spatial-doublers $E^d-E$ vanishes in this limit, as one can
explicitly find from (\ref{asm0d}). 
The energy-momentum relation for the SW action is 
shown in Figure~\ref{fig:disp_SW}. 
The energy at the edge of the Brillouin zone decreases as
$\xi$ increases, which shows the reappearance of the
spatial-doubler.

Since $r_s=1$, the tree-level mass ratio $M_1/M_2$
(Eq.~(\ref{m1m2})) contains no $O((a_sm_0)^n)$ error:
$M_1/M_2 = 1+O((a_tm_0)^2) \inatzero 1$.
The anisotropic SW action has been applied to the simulation 
of charmonium \cite{Umeda:2000ym} and the charmed hadrons
\cite{Matsufuru:2001cp,Harada:2002rf} 
on $\xi \simeq 4$ anisotropic lattices.

\subsection{sD34 action}
We define the sD34 action as
\beq
r_s=c_{\rm SW}^{s}=0  \, , \ \ b_s > 0\, , \ \ d_s > 0.
\eeq
Although the spatial Wilson term is absent ($r_s=0$),  
this action is doubler-free because $d_s > 0$.
The energy splitting $E^d-E$ remains even in the $\atzero$
limit as far as $d_s$ is a constant independent of $\xi$.
Setting $d_s = 1/8$ for this action gives the same 
spatial-doubler mass $m_0^d$ as for the $\xi=1$ SW action
when $\nu=1$. 
The name ``sD34'' is a reminder that the spatial 
$\gamma_iD_i^3$ and $D_i^4$ terms survive in the $\atzero$
limit. 
The sD34 action is similar to the one proposed in 
\cite{Hamber:1983qa,Eguchi:1983xr}, except that those
papers consider the case of the isotropic lattice
$\xi=1$.  
Since the sD34 action has the next-nearest neighbor
interactions such as 
$\psib \del_\mu\De_\mu \psi$ and $\psib \De_\mu^2 \psi$,
this action is more costly to simulate than the SW action
which consists of the nearest neighbor interactions only.

The sD34 action does not generate $O(a_s)$ discretization errors
because of $r_s=c_{\rm SW}^{s}=0$.
In order to remove an $O(a_t)$ error arising from the
temporal Wilson term with $r_t=1$, we take
\beq
\nu = 1 + \frac{1}{2} r_t a_t m_0\, , \ \ c_{\rm SW}^{t} =
\frac{1}{2}\, .
\eeq
This condition is obtained by performing a field redefinition
$\psi_c = \Om_c\psi$, $\psib_c  = \psib\Omb_c$,
$\Omb_c = \Om_c = 1 - {1 \over 4} r_t a_t (D\slash_0 - m_c)$
to the continuum quark action
$\psib_c(x) (D\slash + m_c) \psi_c(x)$
\cite{Alford:1996nx}. 
Since $r_s=0$ the tree-level mass ratio $M_1/M_2$ again
contains no $O((a_sm_0)^n)$ errors: 
$M_1/M_2 = 1+O((a_tm_0)^2) \inatzero  1$.

In the rest of paper we consider the following three choices
of $b_s$ and $d_s$ parameters:
\bea
b_s=\frac{1}{6}\, , \ \ \ \ d_s=\frac{1}{8} \, &:& {\rm ~sD34~~~~~ }, 
\label{eq:sD34} \\
b_s=\frac{1}{8}\, , \ \ \ \ d_s=\frac{1}{8} \, &:& {\rm ~sD34(v)~ },
\label{eq:sD34v}\\
b_s=\frac{1}{2}\, , \ \ \ \ d_s=\frac{1}{4} \, &:& {\rm ~sD34(p)~ }.
\label{eq:sD34p}
\eea
The first choice sD34, where $b_s=1/6$, eliminates an
$O(a_s^2)$ error arising from the $\ga_i \del_i$ term: 
$\ga_i \del_i (1 - \frac{1}{6} a_s^2 \De_i ) =
 \ga_i D_i + O(a_s^4)$.
The second choice sD34(v), where $b_s=1/8$, eliminates 
an $O(a_s^2)$ error in the one-gluon vertex (\ref{gqq}): 
$V_{1,i}^a (q,q',k) = -igt^a\gamma_i + O(a_s^3)$.
The difference between the one-loop results for sD34 and for 
sD34(v) is numerically small as shown in the next section.
With the third choice sD34(p), the hopping terms in the
action are proportional to the projection matrix 
$1\pm \gamma_\mu$:
using the Wilson's projection operator 
$w_\mu ~\equiv~ a_\mu \ga_\mu \del_\mu - \frac{1}{2} a_\mu^2\De_\mu$,
the space-component of the action is given by 
$w_i+\half w_i^2$ ($i=1,2,3$).
Therefore the third choice can reduce simulation costs 
compared to the other two choices \cite{Alford:1996nx}.
At the tree level the sD34 action (\ref{eq:sD34}) is
$O(a_s^2)$-improved, while others contain some $O(a_s^2)$
errors.
Within the current set of operators (\ref{D234}), therefore,
the best available choice to suppress discretization effects
is the sD34 action.

\begin{figure}[ht]
\centerline{
\leavevmode\psfig{file=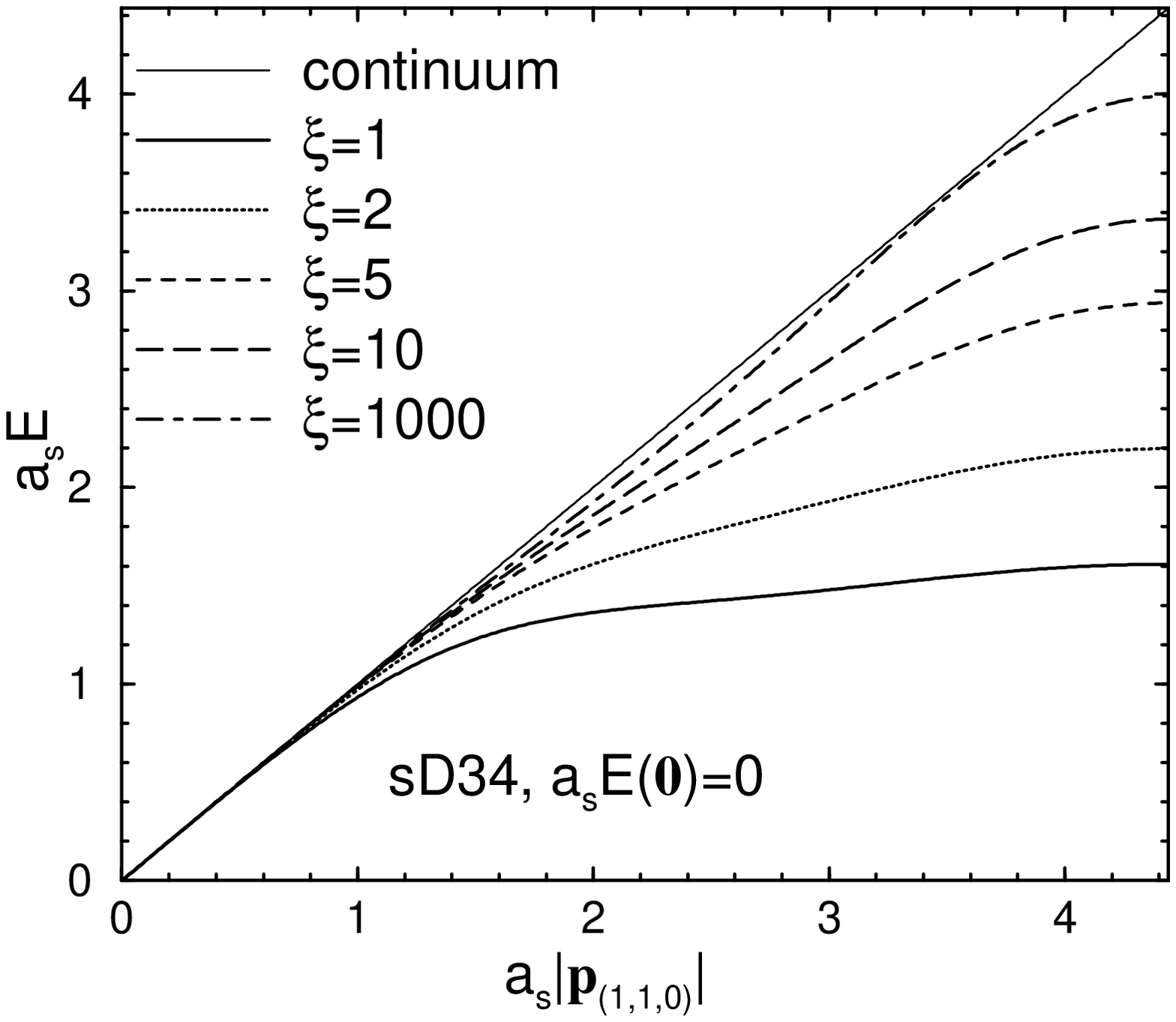,width=8.cm}
\leavevmode\psfig{file=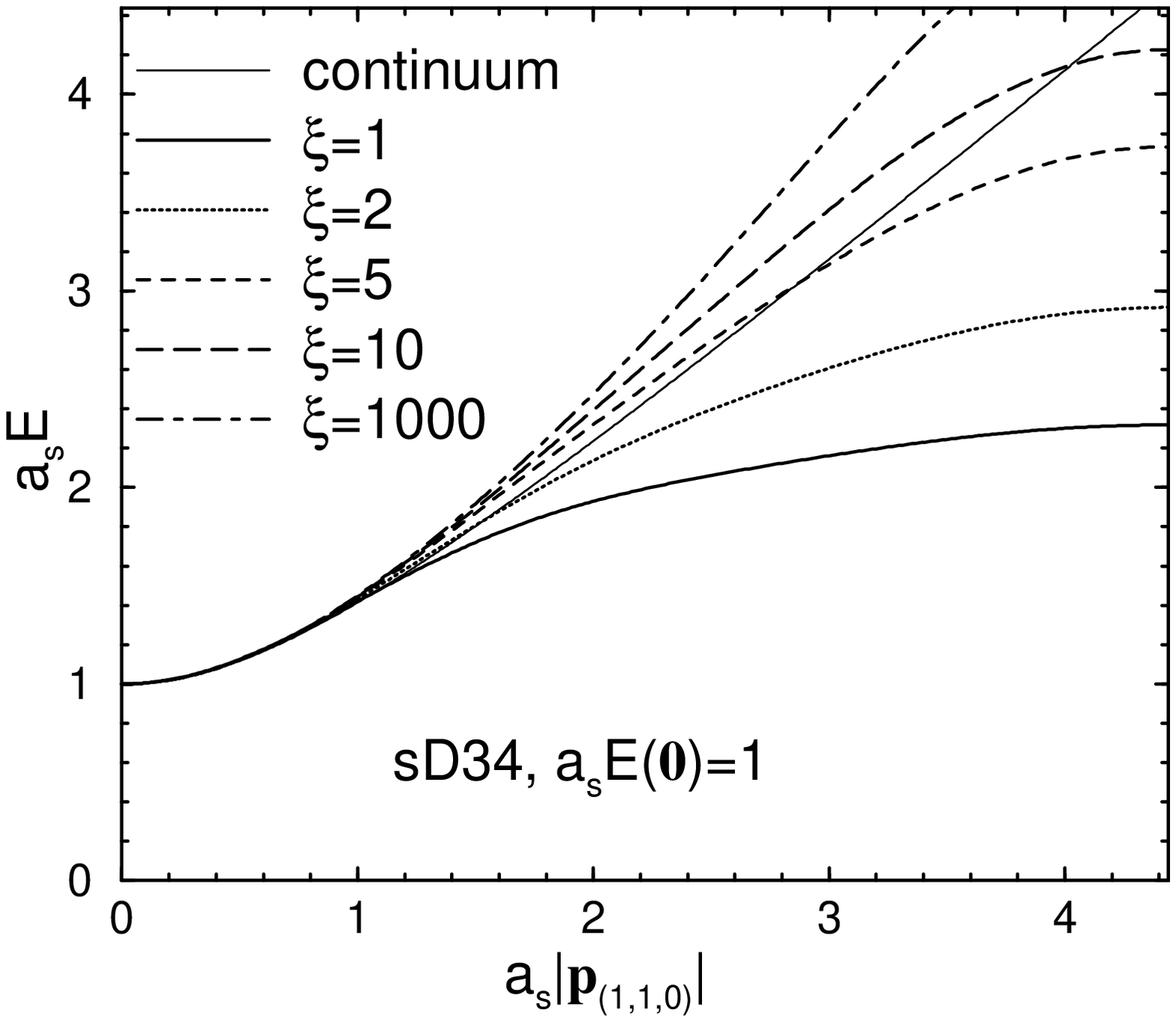,width=8.cm}}
\caption{Energy-momentum relation for the sD34 action. }
\label{fig:disp_sD34}
\end{figure}

\begin{figure}[ht]
\centerline{
\leavevmode\psfig{file=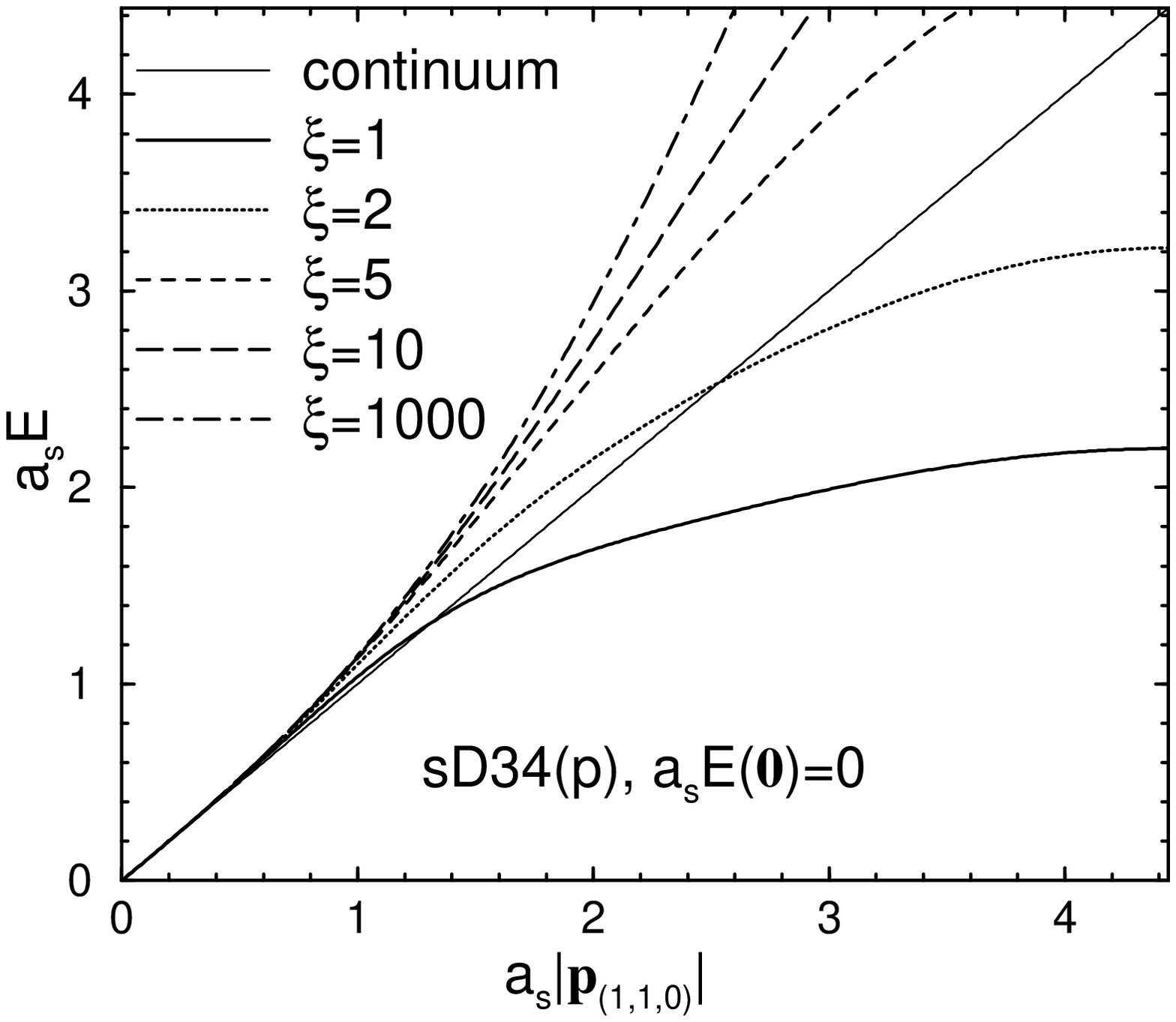,width=8.cm}
\leavevmode\psfig{file=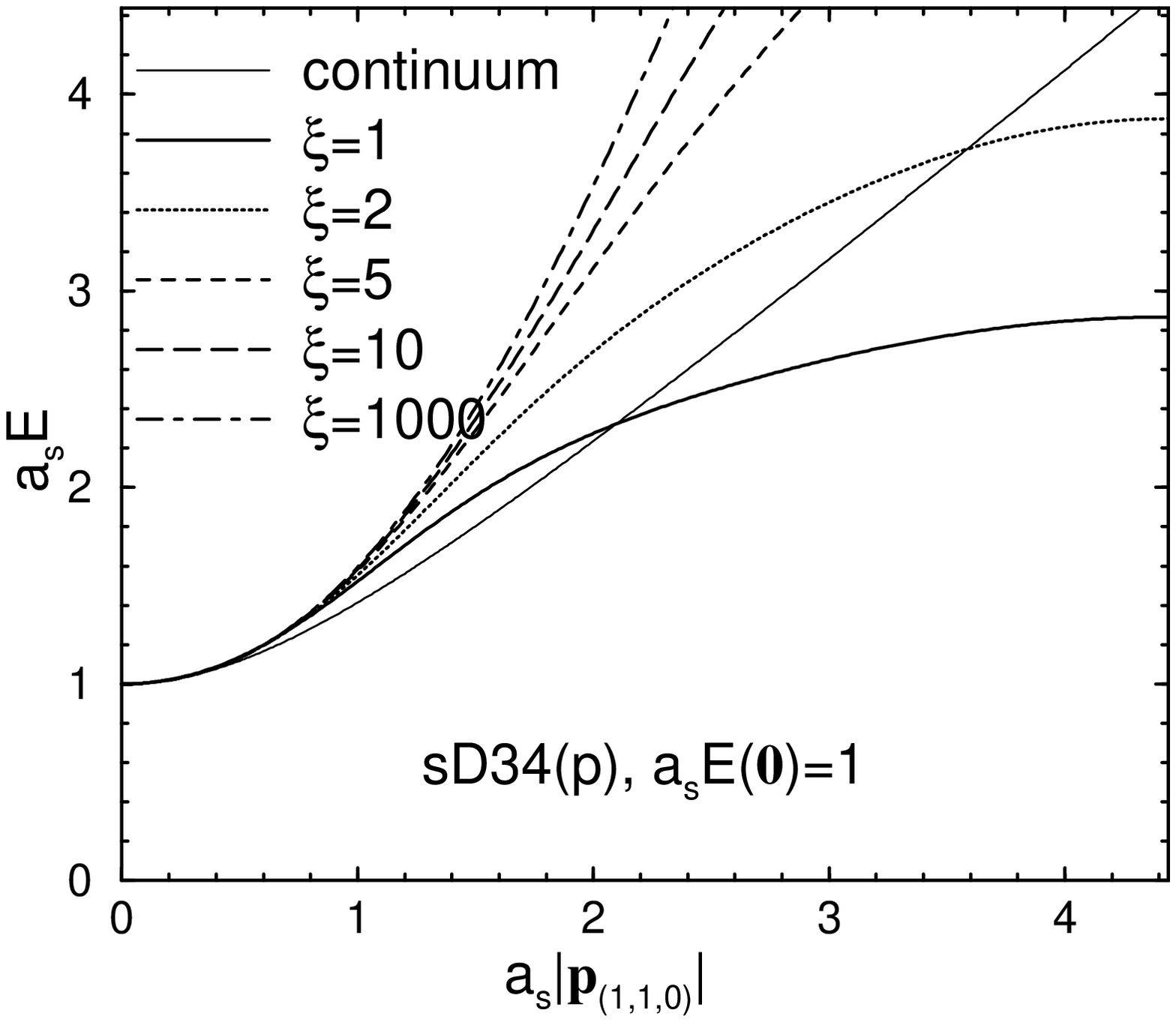,width=8.cm}}
\caption{Energy-momentum relation for the sD34(p) action. }
\label{fig:disp_sD34p}
\end{figure}

The energy-momentum relations for the sD34 and sD34(p)
actions are shown in Figures~\ref{fig:disp_sD34} and
\ref{fig:disp_sD34p} respectively.
For both choices 
the energy at the edge of the Brillouin zone increases as
$\xi$ increases, in contrast to the case of the SW action.
In small $a_s\pbf$ region, the energy-momentum relation for
the sD34 action is quite close to the continuum one because
it has no $O(a_s^2\pbf^2)$ errors. 
Moreover, in large $a_s\pbf$ region near the edge of the
Brillouin zone, it is close to the continuum one too, 
for large values of $\xi$.

To summarize, both the SW action and the sD34 action 
do not generate the $O((a_sm_Q)^n)$ ($n=1,2,\cdots$) errors
at the tree level in the mass ratio $M_1/M_2$.
While the SW action suffers from the spatial doubler for
large values of $\xi$, the sD34 action is doubler-free for
\textit{any} value of $\xi$. 
Both actions can be used for simulations of the charm quark,
if $\xi \simeq 2-4$ and $a_tm_c \ll 1$.
But simulations of the bottom quark keeping $a_tm_b \ll 1$
require $\xi \simeq 5-10$, for which the
anisotropic SW action may be contaminated by the spatial
doublers.

Besides the above actions, two other anisotropic actions 
have been proposed and applied to heavy quark systems:
one is the action with $r_t=1$ and $r_s=\xi$ 
\cite{Klassen:1999fh,Chen:2001ej,Okamoto:2001jb,Mei:2002ip},
and the other is that with  
$r_t=r_s=\xi$ 
\cite{Groote:2000jd,Collins:2001pe,Shigemitsu:2002wh}. 
However, these actions has the spatial Wilson term scaling
as $r_s=\xi$ and therefore generate the $O((a_sm_0)^n)$
errors in the mass ratio $M_1/M_2$ even at the tree level
when $\nu=1$, as discussed before and in \cite{Harada:2001ei}.
For this reason, we do not consider these actions further 
in this paper.

\section{Static limit $\static$ and the Hamiltonian limit
  $\atzero$} 
\label{sec:limitorder}
In this section we discuss the static limit $\static$ of the 
anisotropic fermion actions.
At finite $a_t$, the action always approaches to the usual
static action in the limit of $a_t m_Q\rightarrow\infty$. 
This is shown, \textit{e.g.} by rescaling the fermion field
in (\ref{eq:action}) as 
\beq
\psi(x)=\frac{e^{-a_tM_1\cdot t}}{\sqrt{a_tm_0}}h(x) \, ,
\eeq
and then taking $a_tm_0 \to \infty$.

On the other hand, the action in the limit of $\static$
while keeping the condition $a_t m_Q \ll 1$ can be
different. 
In the $\atzero$ limit, the lattice Dirac operator
(\ref{D234}) becomes
\beq
Q(\atzero) =  m_0 \+ \ga_0 D_0 \+ \nu\sum_i\,  
\ga_i \del_i \, ( 1 - b_s a_i^2 \De_i ) 
\+ \nu d_s a_s^3 \sum_i  \De_i^2,
\label{D234at0}
\eeq
unless $r_s\propto\xi$.
Taking subsequently the static limit, the fermion field
splits as usual into large and small components in the Dirac
representation of the Dirac matrix, and the off-diagonal
terms drop out, the action becomes
\bea
Q(\atzero) &\instatic&  m_0 \+ \ga_0 D_0  
\+ \nu d_s a_s^3 \sum_i  \De_i^2 \, .
\label{D234at0stat}
\eea
We note that the $a_s^3 \De_i^2$ term, proportional to 
$\nu d_s$, remains in the static limit. 
The SW action with $d_s=0$ approaches the usual static
action, but the sD34 action with $d_s > 0$ does not.
This observation is consistent with the static energy
evaluated in the $\atzero$ limit (\ref{Estat}). 
Formally the static limit (\ref{D234at0stat}) can be derived
by applying the Foldy-Wouthuysen-Tani transformation 
\beq
\psi(x) \;\longrightarrow\;
\exp \left[ - \frac{\nu}{2m_0}\sum_i\, 
\ga_i \del_i \, ( 1 - b_s a_i^2 \De_i ) \right] \, \psi(x)
\eeq
to (\ref{D234at0}) and then taking $m_0 \to \infty$. 

Results of the one-loop calculation at $a_t=0$ in the next
section should be consistent with the form
(\ref{D234at0stat}) in the $\static$ limit. 
Suppose that the static action is renormalized as 
\beq
(m_0+\delta_m)  \+ \ga_0 D_0  
\- {1\over 2} \, \delta_r a_s \, \sum_i \Delta_i 
\+ (\nu d_s + \delta_d) a_s^3 \sum_i  \De_i^2 \, ,
\label{D234at0stat_ren}
\eeq
then the mass shift $\delta_m$, the kinetic term
renormalization $\delta_r$, and $\delta_d$ do not depend on 
$m_0$, but may depend on $\nu d_s$ because the static
propagator and vertices contain $\nu d_s$ through
(\ref{D234at0stat}).

\section{One-loop calculation in the Hamiltonian limit}
\label{sec:oneloop}
In this section we present the one-loop calculations for the
anisotropic actions defined in Section~\ref{sec:action}.
We calculate one-loop corrections to the rest mass and the
kinetic mass renormalization factors in the Hamiltonian
limit $\atzero$. 
From the latter we obtain the one-loop correction to the
$\nu$ parameter. 

\subsection{Formalism}
In the one-loop calculation we basically follow the notation
of \cite{Mertens:1997wx} and add some extensions to the case
of the anisotropic lattice.

We write the inverse free quark propagator as
\begin{equation}
  a_t G_0^{-1}(p)= i a_tK\slash (p) + a_tL(p),
\label{freeprop}
\end{equation}
and the self energy as 
\begin{eqnarray}
  a_t\Sigma(p) & = &
  i\sum_\mu \gamma_\mu \sA_\mu(p) \sin(a_\mu p_\mu)  + \sC(p) 
  \nonumber\\
  & \equiv &
  i\sum_\mu \gamma_\mu \sB_\mu(p)                    + \sC(p) \, ,
  \label{self}
\end{eqnarray}
where $p$ is the external momentum.
The inverse full quark propagator is then given by
\begin{equation}
G^{-1}(p)=G_0^{-1}(p)-\Sigma(p).
\end{equation}
Solving 
$G^{-1} (G^{-1})^\dagger = 0$ with $p_0 = i E$,
we obtain the all-orders dispersion relation
\begin{equation}
1+ \mu(\pbf) - \cosh (a_t E) - \sC 
= \sqrt{(1-\sA_0)^2 \sinh^2 (a_t E)-{\textstyle \sum_i
 (a_t K_i-\sA_i\sin(a_s p_i))^2}} \, ,
\label{disp}
\end{equation}
where $\mu(\pbf)$ is given in (\ref{mu}).

Setting $\pzero$ in (\ref{disp}), we obtain the rest mass
$M_1=E(\zbf) \label{M1def}$ as 
\begin{eqnarray}
  \label{M1PT}
  e^{a_tM_1} 
  &=& 1 + a_tm_0 + \sA_0(\on)\sinh (a_tM_1) - \sC(\on) 
  \nonumber\\
  &=& 1 + a_tm_0 -i \sB_0(\on)              - \sC(\on).
\end{eqnarray}
In order to have massless quarks remain massless at the quantum level,
we need a mass subtraction. 
Defining the critical bare mass 
$a_t m_{0c} \equiv  \sC(0,\zbf)$,
we can write 
\begin{equation}
e^{a_tM_1} = 1 + a_tM_0 -i \sB_0(\on) - \sC_\sub(\on),
\end{equation}
where $M_0 = m_0 - m_{0c}$
and 
$\sC_\sub(\on) = \sC(\on) - a_t m_{0c}$.
When $M_0=0$, the rest mass $M_1$ vanishes by construction.
Usually the mass subtraction is done nonperturbatively in
the numerical simulation by defining the critical hopping
parameter. 

In perturbation theory the rest mass is expanded as
\begin{equation}
  M_1 = \sum_{l=0}^\infty g^{2l} M_1^{[l]},
  \label{M1exp}
\end{equation}
in which the tree-level rest mass is 
\begin{equation}
  a_s M_1^{[0]} = \xi\log(1+a_tM_0) \inatzero a_sM_0,
\end{equation}
while the one-loop coefficient is given by
\begin{eqnarray}
  a_s M_1^{[1]} 
  &=& 
  \left(-i \xi\sB_0^{[1]}(\on) - \xi\sC^{[1]}_\sub(\on)\right)
  e^{-a_tM_{1}^{[0]}} 
  \nonumber\\
  &\inatzero& -i \xi\sB_0^{[1]}(\on) - \xi\sC^{[1]}_\sub(\on).
  \label{aM11sub}
\end{eqnarray}
Note that $M_1$ is now normalized by the spatial lattice
spacing $a_s$. 
Before the subtraction the one-loop coefficient is given by
\begin{equation}
  a_s M_{1,\rm nosub}^{[1]} \inatzero 
  -i \xi\sB_0^{[1]}(\on) - \xi\sC^{[1]}(\on) \, .
  \label{aM11nosub}
\end{equation}

Differentiating (\ref{disp}) in terms of $p_1$ twice and then setting
$\pzero$, we obtain the kinetic mass
$M_2=(\partial^2E/\partial p_1^2)^{-1}_{\pzero}$
as 
\begin{equation}
  \label{M2}
  \frac{e^{a_tM_1}-\sA_0(\on)\cosh (a_tM_1)}{\xi^2 a_tM_2}
  = 
  \frac{r_s}{\xi^2} + D(\zbf) 
  + \frac{[\nu/\xi -\sA_1(\on)]^2}{[1-\sA_0(\on)]\sinh (a_tM_1)},
\end{equation}
where
\begin{eqnarray}
  D(\zbf) 
  &=&
  \frac{d^2~}{d(a_sp_1)^2}\left[\sA_0(iE(\pbf),\pbf) \sinh (a_tM_1)
    -\sC(iE(\pbf),\pbf) \right]_{\pzero} 
  \nonumber\\
  &=&
  D_{1s}(\zbf) + \frac{i}{a_sM_2} D_{1t}(\zbf)
  + \frac{i}{a_sM_2}\cdot \frac{\sB_0(\on)}{\xi \tanh (a_tM_1)} 
  \label{D}
\end{eqnarray}
with 
\begin{eqnarray}
  D_{1s}(\zbf) &=&
  \frac{\partial^2~}{\partial (a_sp_1)^2}
  \left[\frac{1}{i}\sB_0 - \sC \right]_{p=(\on)},
  \label{D1s}\\ 
  i D_{1t}(\zbf) &=&
  \frac{\partial~}{\partial (a_sp_0)}
  \left[\sB_0 - i \sC \right]_{p=(\on)}.
  \label{D1t}
\end{eqnarray}
The kinetic mass is expanded as
\begin{equation}
  M_2 = \sum_{l=0}^\infty g^{2l} M_2^{[l]},
  \label{M2exp}
\end{equation}
and the tree level relation becomes
\begin{equation}\label{treeM2}
  \frac{e^{a_tM_1^{[0]}}}{a_t m_2}
  = r_s + \frac{\nu^2}{\sinh (a_tM_1^{[0]})}
\end{equation}
where $M_2^{[0]}=m_2(M_1^{[0]})$.

From (\ref{M2}) and (\ref{treeM2}), we can obtain the 
kinetic mass renormalization factor defined by
\begin{equation}
Z_{M_2} \equiv \frac{M_2}{m_2(M_1)} 
        =  1 + \sum_{l=1}^\infty g^{2l} Z_{M_2}^{[l]}\, . \label{ZM2}
\end{equation}
Here the argument of $m_2$ is the {\it all-orders} rest mass $M_1$.
The one-loop coefficient is given by
\begin{eqnarray}
Z_{M_2}^{[1]} 
        ~=~ &&\frac{2\nu\xi \sA_1^{[1]}(\on)
        -       \nu^2 \sA_0^{[1]}(\on)
        -       D^{[1]}(\zbf)\xi^2\sinh (a_tM_1)} 
           {\nu^2+r_s\sinh (a_tM_1)} \nonumber\\
        && - \sA_0^{[1]}(\on) \cosh (a_tM_1) e^{-a_tM_1}.
        \label{ZM2[1]}
\end{eqnarray}

In the $\atzero$ limit, the tree-level kinetic mass goes to
\begin{equation}
\frac{1}{m_2} \inatzero \frac{\nu^2}{M_1} + R_s a_s \, ,
\end{equation}
where we defined $R_s \equiv r_s/\xi$,
and hence $Z_{M_2}$ goes to
\begin{equation}
Z_{M_2} \inatzero \frac{M_2}{M_1}\nu^2 + R_s a_s M_2 \, .
\label{zm2v}
\end{equation}
Therefore, in this limit, the renormalized $\nu$ parameter and $R_s$
which give $M_1 = M_2$ can be determined from $Z_{M_2}$: 
\begin{eqnarray}
  Z_{M_2} &\inatzero& \nu^2 + R_s a_s M_1 \nonumber\\
          &=& 1 + (2 \nu^{[1]} + R_s^{[1]} a_s M_1^{[0]} )g^2 + \Ogf
\end{eqnarray}
with the one-loop coefficient 
\begin{equation}
Z_{M_2}^{[1]} = 2 \nu^{[1]} + R_s^{[1]} a_s M_1^{[0]} \, .
\label{renorm}
\end{equation}
Here we used $R_s^{[0]} \inatzero 0$.

On the other hand, from (\ref{ZM2[1]}) we obtain in this limit
\begin{equation}
Z_{M_2}^{[1]} \inatzero 
\frac{2}{\nu}\, \xi\sA_1^{[1]}(\on) 
- \frac{2}{i}\,\frac{\xi\sB_0^{[1]}(\on)}{a_sM_1} 
- a_sm_2 \xi D^{[1]}(\zbf) \, ,
\label{zm2at0}
\end{equation}
where 
$\sA_1^{[1]}(\on)=\frac{\partial~\sB_1^{[1]}}{\partial (a_sp_1)}|_{p=(\on)}$.

\subsection{One-loop diagrams}
\begin{figure}[ht]
  \begin{center}
\leavevmode\psfig{file=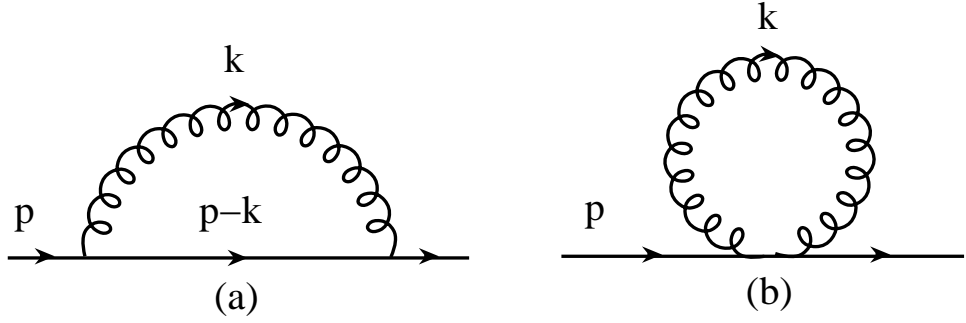,width=15cm}
    \vspace{-2.5cm}
    \caption{Feynman graphs relevant for the one-loop quark self energy.
The left (a) is the regular graph, and the right (b) is the tadpole graph.}
  \label{fig:fgraph}
   \end{center}
\end{figure}

Here we compute one-loop contributions relevant to the rest
mass and the kinetic mass renormalization.  
At the one-loop level, the self-energy is written as 
\beq
\Sigma^{[1]} (p) = \Sigma^\reg(p) + \Sigma^\tad(p) + \Sigma^\TI(p),
\eeq
where the contribution from the regular graph 
Fig.~\ref{fig:fgraph}(a) is denoted by $\Sigma^\reg(p)$, while
the tadpole graph Fig.~\ref{fig:fgraph}(b) gives
$\Sigma^\tad$.
In order to remove the bulk of $\Sigma^\tad$, we apply the 
tadpole improvement \cite{Lepage:1993xa}, which amounts to
$\Sigma^\TI$. 
Feynman rules relevant to our calculations are summarized in 
Appendix \ref{sec:def}.
We use the anisotropic Wilson gluon action 
given by
\beq
S_g = \frac{6}{g^2}\left[\xi \sum_{x,i}(1-P_{0i}(x)) +
\frac{1}{\xi}\sum_{x,i>j}(1-P_{ij}(x)) \right] \, ,
\label{gaugeact}
\eeq
where $P_{0i}(x)$ and $P_{ij}(x)$ are 
the temporal and spatial plaquettes respectively.
In the calculations of $\Sigma^\reg$ and $\Sigma^\tad$,
we adapt the Feynman gauge $\alpha =1$ for the gluon
propagator. 

\subsection{regular graph}
The contribution from the regular graph is 
\bea
a_s\Sigma^\reg(p) 
& = & i\sum_\mu \gamma_\mu \xi\sB_\mu^\reg(p) + \xi\sC^\reg(p) 
\nonumber\\
& = & C_F \int_{-\pi}^{\pi}\frac{d^4 k_\mu}{(2\pi)^4} \,\,
      \gprop \,\,
\frac{i\sum_\mu \gamma_\mu \fBbarmu(p,k) + \fCbar(p,k)}
{\sum_\mu\Kbar_\mu(p-k)^2+\Lbar(p-k)^2} \, ,
\label{selfreg}
\eea
where $\Lbar=a_tL$, $\Kbar=a_tK$ and the gluon momenta are rescaled as 
$a_\mu k_\mu \rightarrow k_\mu$ and $a_\mu \khat_\mu \rightarrow \khat_\mu$,
and 
\bea
\fBbarmu &=& a_t\fB_\mu 
=  2\Kbar_\mu\Xbar_\mu^2 - 
\Kbar_\mu \sum_\rho (\Xbar_\rho^2+\Ybar_\rho^2) +
2\Lbar\Xbar_\mu\Ybar_\mu \, , \\
\fCbar &=& a_t\fC 
= 2\sum_\rho \Kbar_\rho\Xbar_\rho\Ybar_\rho 
-\Lbar \sum_\rho (\Xbar_\rho^2-\Ybar_\rho^2) \, 
\eea
with $\Kbar=\Kbar(p-k)$, $\Lbar=\Lbar(p-k)$, 
$\Xbar_\mu=\Xbar_\mu(2p-k,\pm k)$ and $\Ybar_\mu=\Ybar_\mu(2p-k,\pm k)$
given in Appendix \ref{sec:def}.
Since the vertex from the clover term $\si_{\mu\nu}
F_{\mu\nu}$ in the fermion actions are $O(a_t)$ and
vanishing in the $\atzero$ limit, we omit their contributions.

In the $\atzero$ limit, where
\bea
\fB_0 & \inatzero & (p_0-k_0) \{ 1 - \sum_j (\Xbar_j^2+\Ybar_j^2) \} \, , \\
\fB_i & \inatzero &  2K_i\Xbar_i^2 - 
K_i  \{ 1 + \sum_j (\Xbar_j^2+\Ybar_j^2) \} +
2L\Xbar_i\Ybar_i \, , \\
\fC & \inatzero &  2\sum_j K_j\Xbar_j\Ybar_j
-L \{1+ \sum_j (\Xbar_j^2-\Ybar_j^2)\} \, ,
\eea
we obtain
\bea 
\xi\sB_\mu^\reg(p) & \inatzero & \intk\intinf\frac{dk_0}{2\pi} \,\,
\frac{1}{a_s^2 k_0^2 + \khatbf^2} \,\,
\fB_\mu(p,k) \, S_2(p-k) \, , \label{Breg}\\
\xi\sC^\reg(p) & \inatzero & \intk\intinf\frac{dk_0}{2\pi} \,\,
\frac{1}{a_s^2 k_0^2 + \khatbf^2} \,\,
\fC(p,k) \, S_2(p-k)\, . \label{Creg}
\eea
Here we defined
\beq
\intk ~\equiv~ C_F \int_{-\pi}^{\pi}\frac{d^3 {\bf k}}{(2\pi)^3} \, 
\eeq
and 
\beq
S_2(p-k) ~\equiv~ \frac{1}{(p_0-k_0)^2 + \sum_i K_i(p-k)^2+L(p-k)^2} \, .
\eeq

Differentiating (\ref{Breg}) and (\ref{Creg}) in terms of 
external momenta $p$ and then setting $p=(\on)$, we obtain
the contributions to $M_1^{[1]}$ and $Z_{M_2}^{[1]}$
according to (\ref{aM11sub}) and (\ref{zm2at0}). 
For the evaluation of the loop integrals, we first integrate
over $k_0$ analytically as described in Appendix
\ref{sec:k0int}. 
The remaining integration over $\kbf$ is evaluated
numerically using an adaptive integration routine VEGAS
\cite{Lepage:1980dq}.

Since the rest mass and the kinetic mass are physical quantities,
one-loop corrections to them are infrared-finite. 
Although there are infrared divergences in the partial
derivatives $D_{1s}(\zbf)$ and $D_{1t}(\zbf)$ in the
kinetic mass renormalization, they cancel in the total
derivative $D(\zbf)$. 
In numerical integrations over $\kbf$, we evaluate the total 
derivative directly, rather than evaluate each partial
derivative with subtraction of the infrared divergences.

\subsection{tadpole graph}
Although the calculation of $\Sigma^\tad$ is much simpler than
that of $\Sigma^\reg$, it is worthwhile to show the dependence 
of the results on the mass and on the parameters.
The contribution from the tadpole graph at finite $a_t$
is given by
\bea
a_s\Sigma^\tad(p) 
& = & i\sum_\mu \gamma_\mu \xi\sB_\mu^\tad(p) + \xi\sC^\tad(p) 
\nonumber\\
& = & C_F \int_{-\pi}^{\pi}\frac{d^4 k_\mu}{(2\pi)^4} \gprop \nonumber \\
&& \times \sum_\mu \frac{a_\mu}{a_t} \left[ i\gamma_\mu \left\{\aX \sin(\ap) + 
4\aZ\sin(2\ap)\cos^2(\frac{k_\mu}{2}) \right\}\right. \nonumber \\
&& - \left.\left\{ \aY \cos(\ap) + 4\aW\cos(2\ap)\cos^2(\frac{k_\mu}{2}) 
\right\} \right],
\label{selftad}
\eea
from which we immediately obtain 
\bea
\xi\sB_\mu^\tad(p) & = & C_F\frac{a_\mu}{a_s} \left\{J \, \aX  + 
  8 \, T_\mu\, \aZ\cos(\ap) \right\}\sin(\ap) ,\label{ABtad}\\
\xi\sC^\tad(p) & = & - \,C_F\sum_\mu \frac{a_\mu}{a_s} 
\left\{ J \aY \cos(\ap) + 4\, T_\mu\, \aW\cos(2\ap) \right\},
\label{Ctad}
\eea
where
\bea
J &\equiv& \xi\int_{-\pi}^{\pi}\frac{d^4 k_\mu}{(2\pi)^4} \gprop , \\
T_\mu &\equiv& \xi\int_{-\pi}^{\pi}\frac{d^4 k_\mu}{(2\pi)^4} \gprop 
\, \cos^2(\frac{k_\mu}{2}) .
\eea
In the $\atzero$ limit, $J=T_0=0.2277$ and $T_i=0.1282$.

Tadpole contributions to $M_1^{[1]}$ and $Z_{M_2}^{[1]}$ 
in the $\atzero$ limit are easily calculated from
(\ref{ABtad}) and (\ref{Ctad}). 
The contribution to the rest mass before the subtraction 
(\ref{aM11nosub}) is given by
\beq
a_s M_{1,\rm nosub}^\tad \inatzero 
C_F \sum_i \left\{ J a_sY_i + 4\, T_i\, a_sW_i \right\} ,
\label{m1tad}
\eeq
which depends on $a_sY_i$ and $a_sW_i$ , \textit{i.e.} $\nu
d_s$, but not on the mass. 
It contributes to the critical mass $a_sm_{0c}$ only, so
$a_s M_{1}^\tad=0$ after the subtraction.

The contribution to the kinetic mass renormalization is given by
\beq
Z_{M_2}^\tad \inatzero C_F \left[ 
\frac{2}{\nu} \,(J a_sX_1 + 8\, T_1\, a_sZ_1) \,+\,
a_sm_2 (J a_sY_1 + 16\, T_1\, a_sW_1) \right]  .
\label{zm2tad}
\eeq
We find that a term proportional to $a_sm_2$ appears, which
depends on $\nu d_s$ again. 
This manifest $a_sm_Q$ dependence originates from the $\xi
D_{(1s)}$ term in (\ref{zm2at0}). 
Therefore $Z_{M_2}^\tad $ diverges as $O(a_sm_Q)$ toward the
static limit for the sD34 action with $d_s > 0$, while it is
mass-independent for the SW action with $d_s = 0$. 
Similar mass-dependences are also observed in $Z_{M_2}^\reg$.
We will discuss the $O(a_sm_Q)$ divergence of $Z_{M_2}$ 
in Section~\ref{sec:results}.

\subsection{tadpole improvement}
Tadpole improvement \cite{Lepage:1993xa}
is achieved by replacing the link valuable $U_\mu$ by
$U_\mu/u_\mu$, 
where $u_\mu = \< U_\mu \>$ is the mean link valuable. 
In perturbation theory the contribution from the tadpole
improvement is obtained from the difference between the
inverse free propagator 
$G_0^{-1}$ and the tadpole-improved inverse free propagator
$(G_0^{-1})^\TI$ \cite{Groote:2000jd}. 
In momentum space the latter is given by the former with
replacements 
\bea
\sin(n\ap) \to \sin(n\ap)/u_\mu^n ,~~~
\cos(n\ap) \to \cos(n\ap)/u_\mu^n ,
\eea
where $n=1,2,\cdots$.
The one-loop contribution is then given by
\bea
a_s\Sigma^\TI(p) 
& = & \left( a_s G_0^{-1}(p) - a_s (G_0^{-1})^\TI(p)\right)/g^2 
\nonumber\\
& = & \sum_\mu \frac{a_s}{a_\mu}\, u_\mu^{[1]}
\left[ i\gamma_\mu \left\{2\aX  + 
8\aZ\cos(\ap)\right\}\sin(\ap)\right. \nonumber \\
&& - \left.\left\{ 2\aY \cos(\ap) + 4\aW\cos(2\ap) \right\} \right] ,
\label{selfTI}
\eea
where we expanded 
\beq
u_\mu = 1 + g^2 u_\mu^{[1]} + O(g^4)  ~~~~~(u_\mu^{[1]} < 0).
\eeq
We adopt the mean link in Landau gauge for the definition of $u_\mu$,
which is given by
\beq
u_\mu^{[1]} = - \, \frac{1}{2}\, \frac{a_\mu^2}{a_s^2}\, C_F \, 
J_\mu^{\alpha =0}
\eeq
with
\bea
J_\mu^\alpha &=& \xi\int_{-\pi}^{\pi}\frac{d^4 k_\mu}{(2\pi)^4}\gprop 
\,\left\{ 1 - (1-\alpha) \frac{\frac{a_s^2}{a_\mu^2} \khat_\mu^2}
{\xi^2 \khat_0^2 + \khatbf^2} \right\} \\
&\equiv& J - (1-\alpha)\,\delta J_\mu .
\eea
In the $\atzero$ limit, 
$\sum_{i=1}^3 \delta J_i = \delta J_0 = \frac{1}{2}J$.
We then obtain
\bea
\xi\sB_\mu^\TI(p) & = & -\, C_F\frac{a_\mu}{a_s} J_\mu^{\alpha =0}
\left\{ \aX  + 4 \aZ\cos(\ap) \right\}\sin(\ap) \, ,\label{ABTI}\\
\xi\sC^\TI(p) & = & C_F\sum_\mu \frac{a_\mu}{a_s} J_\mu^{\alpha =0}
\left\{ \aY \cos(\ap) + 2\aW\cos(2\ap) \right\} \, .
\label{CTI}
\eea
Comparing (\ref{ABTI}) and (\ref{CTI}) 
with (\ref{ABtad}) and (\ref{Ctad}), 
we find that $\xi\sB_\mu^\TI$ and
$\xi\sC_\mu^\TI$ 
are also obtained from $\xi\sB_\mu^\tad$ and
$\xi\sC_\mu^\tad$ with replacements
\beq
J \to -\, J_\mu^{\alpha =0}\, , ~~~~~~ 
T_\mu \to - \,\frac{1}{2}\, J_\mu^{\alpha =0} .
\eeq
Contributions to $M_{1,\nosub}^{[1]}$ and $Z_{M_2}^{[1]}$
from the tadpole improvement are given by Eqs.~(\ref{m1tad})
and (\ref{zm2tad}) with the above replacements.
Since $J_i^{\alpha=0}=\frac{5}{6}J$, tadpole contributions
are largely canceled by the tadpole improvement.

\section{One-loop results}
\label{sec:results}

\subsection{Rest mass}
\begin{figure}[th]
  \begin{center}
\leavevmode\psfig{file=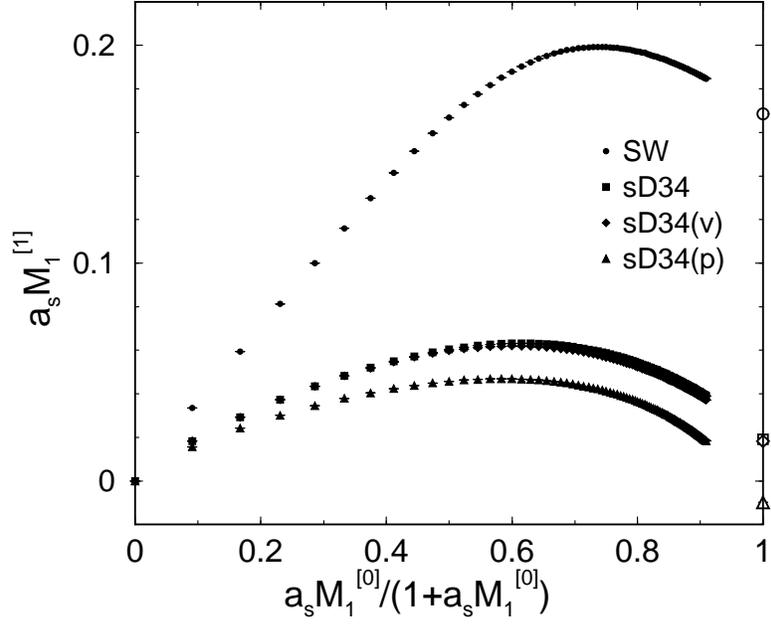,width=10cm}
    \vspace{-.3cm}
    \caption{$a_sM_1^{[1]}$ versus $a_sM_1^{[0]}/(1+a_sM_1^{[0]})$ 
for the SW action and the sD34 actions. 
The values in the static limit are denoted by open symbols.}
  \label{fig:am_aM11}
   \end{center}
\end{figure}

Now we present the results of our one-loop calculations
in the $\atzero$ limit.
The one-loop correction to the rest mass $a_sM_1^{[1]}$ is
plotted as a function of $a_sM_1^{[0]}/(1+a_sM_1^{[0]})$ in
Figure~\ref{fig:am_aM11}, 
and numerical values of $a_sM_1^{[1]}$ and $a_sm_{0c}^{[1]}$
are given in Table~\ref{tab:m1}.
As shown in the figure, $a_sM_1^{[1]}$ for all the actions 
increase from the massless limit and reach their maximum
values around $a_sM_1^{[0]}=1$--3, then 
decrease to the static values represented by open symbols.
Fitting our results in the small mass region 
($a_sM_1^{[1]} \ll 1$), we confirmed that the one-loop
corrections are consistent with the mass singularity
\beq
M_1^{[1]} \sim
- C_F\frac{3}{16\pi^2}M_1^{[0]}\log (a_sM_1^{[0]})^2.
\eeq

The results for $a_sM_1^{[1]}$ in the static limit 
$a_sM_1^{[0]} \to \infty$ depend on the action,
since the reduced static action (\ref{D234at0stat}) includes
the $a_s^3 \De_i^2$ term proportional to $\nu d_s$. 
This situation is in contrast to the case of finite $a_t$
calculations in \cite{Groote:2000jd,Mertens:1997wx},
where $a_sM_1^{[1]}$ goes to a universal value.
In the finite $a_t$ case, $a_sM_1^{[1]}$ does not depend on
$\nu d_s$ in the $a_t\static$ limit, because the static
action always gives the Wilson line. 

\subsection{Kinetic mass renormalization}
\begin{figure}[hbt]
  \begin{center}
\leavevmode\psfig{file=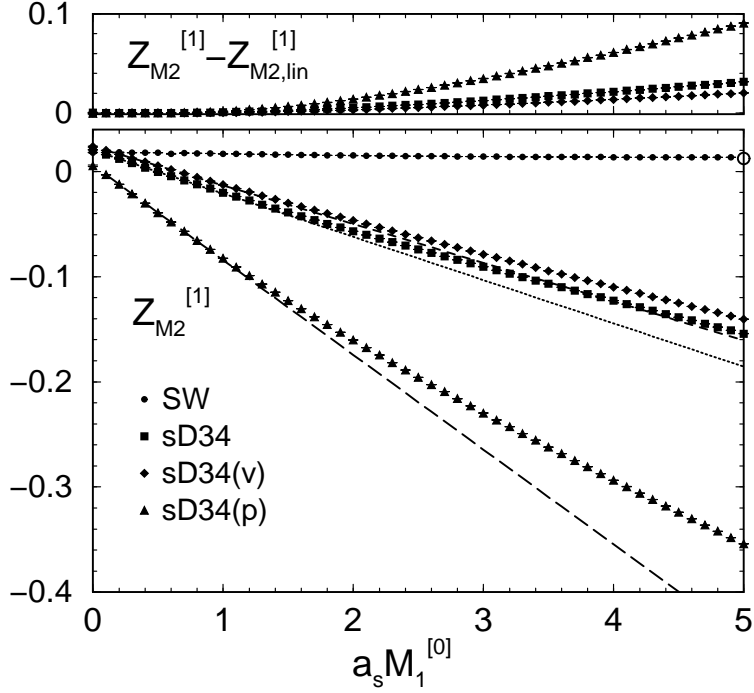,width=10cm}
    \vspace{-.3cm}
    \caption{The lower figure shows $Z_{M_2}^{[1]}$ versus $a_sM_1^{[0]}$ for the SW action 
and the sD34 actions. The value in the static limit for the SW action
is denoted by open circle. Lines are the linear approximations to the
results for the sD34 actions ($Z_{M_2,\rm lin}^{[1]}$) as explained in the text.
The upper figure shows the difference $Z_{M_2}^{[1]}-Z_{M_2,\rm lin}^{[1]}$ 
versus $a_sM_1^{[0]}$ for the sD34 actions.
}
  \label{fig:zm2all}
   \end{center}
\end{figure}

The one-loop correction to the kinetic mass renormalization
$Z_{M_2}$ is related to the speed of light renormalization
$\nu$ according to (\ref{renorm}).
The study of the $a_sm_Q$ dependence of $\nu$ at the
one-loop level is a main purpose of this paper.
The results of $Z_{M_2}^{[1]}$ are shown in
Figure~\ref{fig:zm2all}, and their numerical values are
given in Table~\ref{tab:zm2}. 

First, we focus on the result for the SW action, which
becomes the naive quark action in the $\atzero$ limit as the
Wilson term and the clover term vanish.
From Figure~\ref{fig:zm2all} (lower panel), we find that the
mass dependence of $Z_{M_2}^{[1]}$ (filled circle) is very
weak, and $Z_{M_2}^{[1]}$ stays constant in the infinite
mass limit.
A difference between the value in the static limit and that
in the massless limit is 
$Z_{M_2}^{[1]}(\infty )-Z_{M_2}^{[1]}(0) = -0.006$.
This is only 6\% of the same difference for the isotropic SW
action $-0.10$ \cite{Mertens:1997wx}. 
The result implies that mass dependent discretization errors
of order $g^2(a_sm_Q)^n$ for $Z_{M_2}^{[1]}$
are small on the anisotropic lattice.
The same conclusion holds for any action which becomes the
naive quark action in the $\atzero$ limit.
For instance, the action with $r_s=0$ and $d_s=d/\xi$, where
$d$ is a constant independent of $\xi$, belongs to this
class. 
However, we remark that such actions suffer from the spatial
doublers for large values of $\xi$, as mentioned in
Section~\ref{sec:action}. 

Next, we consider the results for the sD34 actions, which
are doubler-free even in the $\atzero$ limit.
As shown in Figure~\ref{fig:zm2all} (lower panel),
$Z_{M_2}^{[1]}$ for the sD34 actions monotonically decreases
as the mass increases, and diverges as $O(a_sm_Q)$ toward
the static limit.

The $O(a_sm_Q)$ divergence of $Z_{M_2}^{[1]}$ is due to
the finiteness of $\xi D_{(1s)}$ in the static limit
multiplied by $a_sm_2$ in (\ref{zm2at0}).
The appearance of this manifest $a_sm_Q$ dependence, which
is proportional to $\nu d_s$, can be explained as follows.
Using $Z_{M_2}^{[1]}$, the kinetic term renormalization
$\delta_r$ for the static action (\ref{D234at0stat_ren})
is given by 
\beq
\delta_r = \lim_{\static} \left( \frac{1}{a_sM_2} - \frac{1}{a_sm_2} \right)
         = - \lim_{\static} \, g^2 \frac{Z_{M_2}^{[1]}}{a_sm_2} + \Ogf .
\eeq
Because $\delta_r$ is a constant independent of the mass,
${Z_{M_2}^{[1]}}$ diverges as $O(a_sm_Q)$ in the large mass
limit.

\begin{figure}[th]
  \begin{center}
\leavevmode\psfig{file=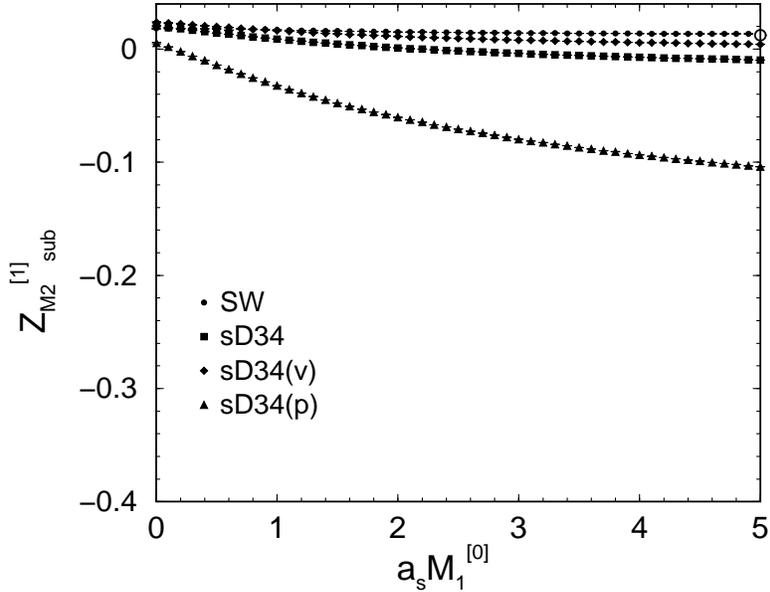,width=10cm}
    \vspace{-.3cm}
    \caption{$Z_{M_2,\rm sub}^{[1]}$ for the sD34 actions
together with $Z_{M_2,\rm sub}^{[1]}=Z_{M_2}^{[1]}$ 
for the SW action. }
  \label{fig:zm2sub}
   \end{center}
\end{figure}

However, we note that this kind of $O(a_sm_Q)$ divergence is nothing to
do with the discretization error increasing as $a_s m_Q$ but
a renormalization of the reduced static action
(\ref{D234at0stat_ren}).
In order to isolate such an $O(g^2 a_sm_Q)$ effect, we
consider a subtracted $Z_{M_2}^{[1]}$ defined through
\beq
\label{eq:Z_M2_sub}
Z_{M_2,\rm sub}^{[1]} = Z_{M_2}^{[1]} + \delta_r^{[1]}\ a_sm_2
\eeq
as a measure of the remaining $O(g^2(a_sm_Q)^n)~(n\ge 2)$
errors.
After the subtraction of the manifest $a_s m_Q$ dependence,
$Z_{M_2,\rm sub}^{[1]}$ for the sD34 actions converges to
a finite value in the static limit as shown in
Figure~\ref{fig:zm2sub}.
We also find that the mass dependence of 
$Z_{M_2,\rm sub}^{[1]}$ for the sD34 and sD34(v) actions is
as small as that for the SW action. 
Note that $Z_{M_2,\rm sub}^{[1]}=Z_{M_2}^{[1]}$ for the SW action 
because of $\delta_r=0$.

Another way to discuss the remaining 
$O(g^2(a_sm_Q)^n)~(n\ge 2)$ errors for the sD34 actions is
to assess their linearity in the mass parameter. 
Since $Z_{M_2}^{[1]}$ for the sD34 actions seems like a
linear function of $a_s M_1^{[0]}$ effectively as shown in
Figure~\ref{fig:zm2all} (lower panel), we attempt a linear
fit using the data for $a_sM_1^{[0]} \le 0.5$. 
The fitting lines
$Z_{M_2,\rm lin}^{[1]}=Z_{M_2}^{[1]}(0) 
 + c_r^{[1]}\times \, a_s M_1^{[0]}$
shown by dashed or dotted lines approximate $Z_{M_2}^{[1]}$
very well from the small mass region $a_sM_1^{[0]} \ll 1$
to a relatively large mass regime $a_sM_1^{[0]} \sim 1$.
The difference $Z_{M_2}^{[1]}-Z_{M_2,\rm lin}^{[1]}$ is
plotted in the upper panel of Figure~\ref{fig:zm2all}.
We find that the difference is less than or about 0.005
(0.01) at $a_sM_1^{[0]} =1$ (3) for the sD34 and sD34(v)
actions, and slightly larger for the sD34(p) action.
Since the (renormalized) coupling constant is 
$g^2=4\pi \alpha_s\sim 2$ in current simulations,
the difference from the linearity
$g^2 (Z_{M_2}^{[1]}-Z_{M_2,\rm lin}^{[1]})$ is small
compared to the tree-level value $Z_{M_2}^{[0]}=1$. 
It indicates that $O(g^2(a_sm_Q)^n)~(n\ge 2)$ errors 
are suppressed on the anisotropic lattice, and
$Z_{M_2}^{[1]}$ for the sD34 actions can be well
approximated by a linear ansatz; 
$Z_{M_2}^{[1]}~\approx~Z_{M_2,\rm lin}^{[1]}$.

If one would like to avoid the appearance of the
renormalization scaling as $a_s m_Q$, it is possible to tune
the spatial Wilson term as $R_s^{[1]} = - \delta_r^{[1]}$ such
that the second term in (\ref{eq:Z_M2_sub}) vanishes,
and then the one-loop coefficient of the speed-of-light
renormalization is given by $\nu^{[1]} = Z_{M_2,\rm sub}^{[1]}/2$.
Since the remaining $O((a_s m_Q)^n)$ correction for $\nu^{[1]}$ is
small and does not diverge
as a function of $a_s m_Q$ as shown in
Figure~\ref{fig:zm2sub}, it essentially solves the problem
of large radiative correction in the anisotropic lattice
actions for heavy quark.
It also suggests that if one can nonperturbatively 
tune the Wilson term in the static limit, e.g. by 
adjusting $r_s$ until the $O(a_s m_Q)$ divergence of 
$Z_{M_2}$ for mesons goes away, 
the above
cancellation of the $a_s m_Q$ error can be implemented
nonperturbatively.

\section{Conclusions}
\label{sec:concl}

In this paper we discuss on the issue whether the
discretization error scales as $(a_s m_Q)^n$ when the heavy
quark action is discretized on an anisotropic lattice for
which the temporal lattice spacing $a_t$ is very small in
order to keep the condition $a_t m_Q \ll 1$ while the
spatial lattice spacing $a_s$ is relatively large and 
$a_s m_Q$ can be order one.
Our naive expectation is that the discretization error does
not behave as $a_s m_Q$ for the heavy-light mesons (or
baryons) at rest, since momentum scale flowing into the
spatial direction is of order of the QCD scale
$\Lambda_{\mathrm{QCD}}$ rather than the heavy quark mass
scale $m_Q$.
Even at the quantum level the maximum (virtual) momentum
flowing into the spatial direction is $\pi/a_s$, and the
discretization error coming from the spatial derivative
cannot pick up the large heavy quark mass.

Through the one-loop calculations of the kinetic mass
renormalization for a class of lattice fermion actions, we
found that our expectation is indeed the case.
For the sD34 actions there is a piece which behaves as 
$a_s m_Q$ in the one-loop coefficient of the kinetic mass
renormalization, but it originates from the renormalization
of the spatial Wilson term, which remains even in the static
limit, and thus does not come from the discretization of the
spatial derivative.
It implies that if one can nonperturbatively tune the
spatial Wilson term (the parameter $r_s$) such that it
vanishes in the static limit, the unwanted behavior 
$a_s m_Q$ can be removed from the speed-of-light
renormalization.
Although there is a possibility that the unwanted
discretization error scaling as $a_s m_Q$ exists in some
other quantities, it is unlikely from our
considerations.

The anisotropic lattice thus remains as a promising approach
to treat heavy quarks on the lattice.
As in the usual relativistic approach, the theory is
renormalizable and the number of necessary terms in the
action is limited.
It also opens a possibility to tune the parameters in the action
nonperturbatively for heavy quarks.

\section*{Acknowledgments}

We thank Tetsuya~Onogi, Shinichi~Tominaga and
Norikazu~Yamada for useful discussions. 
We also thank Andreas~Kronfeld for carefully reading the
manuscript. 
This work is supported in part by Grants-in-Aid of the
Ministry of Education under the contract No.~14540289.
M.O. is also supported by the JSPS.

\appendix
\section{Definitions and Feynman rules}
\label{sec:def}

The lattice covariant derivatives are defined by
\bea
 \del_\mu \psi(x) &\equiv& 
 {1\over 2a_\mu}\, \biggl[ U_\mu(x) \psi(x+\mu) - U_{-\mu}(x) 
           \psi(x-\mu)\biggr] \, ,\\ 
 \De_\mu \psi(x)  &\equiv& 
 {1\over a_\mu^2} \, \biggl[ U_\mu(x) \psi(x+\mu) + U_{-\mu}(x) \psi(x-\mu)
                                       -2 \psi(x) \biggr] \, ,\\ 
 \del_\mu\De_\mu \psi(x) &\equiv& 
 {1\over 2a_\mu^3}\, \biggl[ U_\mu(x) U_\mu(x+\mu) \psi(x+2\mu) 
- U_{-\mu}(x) U_{-\mu}(x-\mu) \psi(x-2\mu) \nonumber\\ 
&& -2 U_\mu(x) \psi(x+\mu) + 2 U_{-\mu}(x) \psi(x-\mu)\biggr] \, ,\\ 
 \De_\mu^2 \psi(x)  &\equiv&
 {1\over a_\mu^4} \, \biggl[ U_\mu(x) U_\mu(x+\mu) \psi(x+2\mu) 
+ U_{-\mu}(x) U_{-\mu}(x-\mu) \psi(x-2\mu) \nonumber\\ 
&&-4 U_\mu(x) \psi(x+\mu) -4 U_{-\mu}(x) \psi(x-\mu)
                                       +6 \psi(x) \biggr] \, .
\eea
We also define the lattice momenta 
\bea
a_\mu \pbar_\mu &\equiv& \sin(a_\mu p_\mu) , \\
a_\mu \phat_\mu &\equiv& 2 \sin(a_\mu p_\mu/2) .
\eea

Feynman rules for our anisotropic actions can be derived in  
usual way. The gluon propagator 
with Feynman gauge 
is given by
\beq
D_{\mu\nu}^{ab}(k) = \frac{\delta^{ab}\delta_{\mu\nu}}{\khat^2} .
\eeq
The quark propagator is
\beq
G_0(p)= \frac{1}{i \sum_\mu\gamma_\mu K_\mu(p) + L(p)} ,
\eeq
where 
\bea
        K_0(p) & = & \pbar_0 \inatzero p_0 ,          \\
        K_i(p) & = & \nu \pbar_i (1+ b_s a_i^2 \phat_i^2 ), \label{Kmu} 
\eea
and 
\bea
        L(p)  & = & m_0 + {1\over 2} a_t \sum_\mu r_\mu \phat_\mu^2 
                    + \nu d_s  \sum_i a_i^3 \phat_i^4 \\
              &\inatzero& m_0 + \nu d_s  \sum_i a_i^3 \phat_i^4 
\eea
for our quark actions with Eq.~(\ref{sD234}).

The one-gluon vertex with the incoming quark momentum $q$, 
the outgoing quark momentum $q'$ and the incoming gluon momentum $k=q'-q$ 
is given by
\beq
V_{1,\mu}^a (q,q',k) = -i g t^a \left[ 
\gamma_\mu \Xbar_\mu(q+q',k) - i \Ybar_\mu(q+q',k)
\right] ,
\label{gqq}
\eeq
where
\bea
\Xbar_\mu(q+q',k) & = & 
2\aX \cos \left( \frac{a_\mu q_\mu+a_\mu q'_\mu}{2}\right)  
+ 4\aZ \cos \left( a_\mu q_\mu+a_\mu q'_\mu\right)
\cos \left(\frac{a_\mu k_\mu}{2}\right) , \\
\Ybar_\mu(q+q',k) & = &
2\aY \sin \left( \frac{a_\mu q_\mu+a_\mu q'_\mu}{2}\right)  
+ 4\aW \sin \left( a_\mu q_\mu+a_\mu q'_\mu\right)
\cos \left(\frac{a_\mu k_\mu}{2}\right) ,
\eea
and 
\bea
\aX & = & \frac{1}{2}\,\nu_\mu + \nu_\mu b_\mu ,   \\
\aY & = & \frac{1}{2}\,r_\mu\frac{a_0}{a_\mu} + 4\nu_\mu d_\mu ,    \\
\aZ & = & -\frac{1}{2}\,\nu_\mu b_\mu ,      \\
\aW & = & -\nu_\mu d_\mu .  
\eea
The $t^a$ are generators of color SU(3).
We ignore the one-gluon vertex arising from the clover terms
because such a vertex becomes irrelevant in the $\atzero$ limit.

Finally the two-gluon vertex with 
the incoming gluon momenta $k$ and $k'$ ($k+k'=q'-q$)
is given by
\bea
&&V_{2,\mu\mu}^{ab} (q,q',k,k') =  2 a_\mu g^2 (t^a t^b) \times\nonumber\\
&& \left[ 
i \gamma_\mu \left\{
\aX \sin \left( \frac{a_\mu q_\mu+a_\mu q'_\mu}{2}\right)  
+ 4\aZ \sin \left( a_\mu q_\mu+a_\mu q'_\mu\right)
\cos\left(\frac{a_\mu k_\mu}{2}\right)\cos\left(\frac{a_\mu k'_\mu}{2}\right)  
\right\} \right.\nonumber\\
&& \left. - \left\{ 
\aY \cos \left( \frac{a_\mu q_\mu+a_\mu q'_\mu}{2}\right)  
+ 4\aW \cos \left( a_\mu q_\mu+a_\mu q'_\mu\right)
\cos\left(\frac{a_\mu k_\mu}{2}\right)\cos\left(\frac{a_\mu k'_\mu}{2}\right)
\right\}
\right] .
\label{ggqq}
\eea
Here we omit terms that vanish by symmetrizing between two gluons 
and that arise from the clover terms, which are unnecessary
in the calculation of the tadpole graph.

\section{$k_0$-integrations}
\label{sec:k0int}
In this Appendix we summarize some formula on the $k_0$-integrations,
which are needed for the calculation of the regular graph.
We use the following results for one-dimensional integrations:
\bea
I_1 &\equiv& \intinf dx\, \ig \, \frac{c}{\iqd} \nonumber\\
&=& \pi c \left\{ 
\frac{1}{\sqrt{ab}}\, \frac{1}{-g^2(e-\sqrt{a/b})^2 +f^2} + 
\frac{1}{a-b(e+f/g)^2} \,\frac{1}{fg}
\right\} ,\\
I_2&\equiv& \intinf dx\, \ig \, \frac{c(\iqn)}{\iqd} \nonumber\\
&=& i\pi c \left\{ 
\frac{1}{\sqrt{ab}}\, \frac{e-\sqrt{a/b}}{-g^2(e-\sqrt{a/b})^2 +f^2} -
\frac{1}{a-b(e+f/g)^2} \,\frac{1}{g^2}
\right\} , \\
I_3 &\equiv& \intinf dx\, \ig \, \frac{c}{(\iqd)^2} 
~=~ - \frac{1}{2f} \frac{\p I_1}{\p f} , \\
I_4 &\equiv& \intinf dx\, \ig \, \frac{c(\iqn)^2}{(\iqd)^2} 
~=~ - \frac{1}{2g} \frac{\p I_1}{\p g} , \\
I_5 &\equiv& \intinf dx\, \ig \, \frac{c(\iqn)}{(\iqd)^2} 
~=~ - \frac{1}{2f} \frac{\p I_2}{\p f} , \\
I_6 &\equiv& \intinf dx\, \ig \, \frac{c}{(\iqd)^3} 
~=~ - \frac{1}{4f} \frac{\p I_3}{\p f} , \\
I_7 &\equiv& \intinf dx\, \ig \, \frac{c(\iqn)}{(\iqd)^3} 
~=~ - \frac{1}{4f} \frac{\p I_5}{\p f} ,
\eea
where $e < f/g$ is assumed. 
These integrations are calculated by hand using the residue theorem,
and checked by Mathematica. 

In the calculation of the regular graph, we assign 
\beq
x \to k_0 \, , \ \ \ a \to |\khatbf|^2\, , \ \ \ b \to a_s^2\, , \ \ \ g
\to 1\, , \ \ \ e \to M_1\, , \ \ \  f \to E(\kbf)\, ,
\eeq
where
\beq
E(\kbf) \equiv \sqrt{\nu^2\sum_i \kbar_i^2 (1+b_sa_i^2\khat_i^2)^2
+ (m_0+\nu d_s \sum_i a_i^3 \khat_i^4)^2}\, .
\eeq
The overall factors $c$ depend on the spatial momentum $\kbf$.
Using integrations $I_1$--$I_7$ with above assignments,
relevant contributions from the regular graph are given by
\bea
\xi\sB_0^\reg(\on) & = & \intk \frac{1}{2\pi} I_{2-B0} \, ,\\
\xi\sC^\reg(\on)   & = & \intk \frac{1}{2\pi} I_{1-C} \, ,\\
\xi\sA_1^\reg(\on) & = &  
\intk \frac{1}{2\pi} (I_{1-A1} + I_{3-A1})\, ,\\
\xi D_{1s}^\reg(\zbf) & = & \intk \frac{1}{2\pi} \left[
\frac{1}{i}(I_{2-Ds}+I_{5-Ds}+I_{7-Ds})-(I_{1-Ds}+I_{3-Ds}+I_{6-Ds})
\right] \, ,\\
i\,\xi D_{1t}^\reg(\zbf) & = & \intk \frac{1}{2\pi} 
(I_{1-Dt}+I_{4-Dt}-iI_{5-Dt}) \, .
\eea

\clearpage 

\begin{table}[ht]
\begin{center}
\begin{tabular}{c|rrrr}
\hline\hline
\multicolumn{5}{c}{$a_s M_1^{[1]}$} \\\hline
$a_s M_1^{[0]}$& \multicolumn{1}{c}{SW} & \multicolumn{1}{c}{sD34} &
\multicolumn{1}{c}{sD34(v)} & \multicolumn{1}{c}{sD34(p)} \\\hline
$  0.0$ & $  0.000000(00)$ & $  0.000000(00)$ & $  0.000000(00)$ & $  0.000000(00)$ \\ 
$  0.1$ & $  0.033586(11)$ & $  0.018299(19)$ & $  0.018360(21)$ & $  0.015615(23)$ \\ 
$  0.2$ & $  0.059382(20)$ & $  0.029235(21)$ & $  0.029306(21)$ & $  0.024206(23)$ \\ 
$  0.3$ & $  0.081252(22)$ & $  0.037292(24)$ & $  0.037354(23)$ & $  0.030177(41)$ \\ 
$  0.4$ & $  0.100023(23)$ & $  0.043466(28)$ & $  0.043405(21)$ & $  0.034594(25)$ \\ 
$  0.5$ & $  0.116008(27)$ & $  0.048228(29)$ & $  0.048196(25)$ & $  0.037909(25)$ \\ 
$  0.6$ & $  0.129786(19)$ & $  0.051966(24)$ & $  0.051762(25)$ & $  0.040428(32)$ \\ 
$  0.7$ & $  0.141467(36)$ & $  0.054948(34)$ & $  0.054572(31)$ & $  0.042396(31)$ \\ 
$  0.8$ & $  0.151341(22)$ & $  0.057183(24)$ & $  0.056763(22)$ & $  0.043836(33)$ \\ 
$  0.9$ & $  0.159679(29)$ & $  0.059029(36)$ & $  0.058465(29)$ & $  0.044916(49)$ \\ 
$  1.0$ & $  0.166758(21)$ & $  0.060338(33)$ & $  0.059687(24)$ & $  0.045724(34)$ \\ 
$  2.0$ & $  0.196227(24)$ & $  0.062578(35)$ & $  0.060997(28)$ & $  0.045709(32)$ \\ 
$  3.0$ & $  0.199166(39)$ & $  0.058705(39)$ & $  0.056713(28)$ & $  0.041241(34)$ \\ 
$  4.0$ & $  0.197161(26)$ & $  0.054263(23)$ & $  0.052280(29)$ & $  0.036472(30)$ \\ 
$  5.0$ & $  0.194441(44)$ & $  0.050519(28)$ & $  0.048617(55)$ & $  0.032255(53)$ \\ 
$ 10.0$ & $  0.184712(29)$ & $  0.038993(46)$ & $  0.037121(28)$ & $  0.018373(37)$ \\ 
$\infty$& $  0.168490(26)$ & $  0.019045(21)$ & $  0.018249(29)$ & $ -0.009803(27)$ \\
\hline\hline
 \multicolumn{5}{c}{$a_s m_{0c}^{[1]}$} \\\hline
   -    & $  0.000000(00)$ & $ -0.060828(02)$ & $ -0.061672(02)$ & $  0.001431(03)$ \\
\hline\hline
\end{tabular}
\caption{Numerical values of $a_sM_1^{[1]}$ for various values of 
$a_s M_1^{[0]}$, 
and $a_s m_{0c}^{[1]}$ for the SW action and the sD34 actions.}
\label{tab:m1}
\end{center}
\end{table}

\begin{table}[ht]
\begin{center}
\begin{tabular}{c|rrrr}
\hline\hline
\multicolumn{5}{c}{$Z_{M_2}^{[1]}$} \\\hline
$a_s M_1^{[0]}$& \multicolumn{1}{c}{SW} & \multicolumn{1}{c}{sD34} &
\multicolumn{1}{c}{sD34(v)} & \multicolumn{1}{c}{sD34(p)} \\\hline
$  0.0$ & $  0.01810(35)\,\ $ & $0.02061(30)\,\ $ & $0.02369(49)\,\ $ & $0.00549(59)\,\ $ \\ 
$  0.1$ & $  0.01812(19)\,\ $ & $0.016066(63)   $ & $0.020152(61)$ & $ -0.00313(11)\,\ $ \\ 
$  0.2$ & $  0.018236(53)$ & $  0.012019(46)$ & $  0.016416(46)$ & $ -0.012140(85)$ \\ 
$  0.3$ & $  0.017991(51)$ & $  0.007856(48)$ & $  0.012592(45)$ & $ -0.021208(59)$ \\ 
$  0.4$ & $  0.017941(49)$ & $  0.003750(39)$ & $  0.009003(38)$ & $ -0.030040(78)$ \\ 
$  0.5$ & $  0.017801(43)$ & $ -0.000349(34)$ & $  0.005406(47)$ & $ -0.039223(56)$ \\ 
$  0.6$ & $  0.016708(38)$ & $ -0.004308(39)$ & $  0.001676(33)$ & $ -0.047996(50)$ \\ 
$  0.7$ & $  0.017507(32)$ & $ -0.008387(32)$ & $ -0.001963(32)$ & $ -0.056888(53)$ \\ 
$  0.8$ & $  0.017326(64)$ & $ -0.012299(84)$ & $ -0.005543(28)$ & $ -0.065500(50)$ \\ 
$  0.9$ & $  0.017142(47)$ & $ -0.016238(28)$ & $ -0.009117(28)$ & $ -0.074070(46)$ \\ 
$  1.0$ & $  0.016862(36)$ & $ -0.020149(28)$ & $ -0.012661(31)$ & $ -0.082456(55)$ \\ 
$  2.0$ & $  0.015176(35)$ & $ -0.056742(34)$ & $ -0.046665(28)$ & $ -0.160259(59)$ \\ 
$  3.0$ & $  0.014246(21)$ & $ -0.090607(29)$ & $ -0.078766(30)$ & $ -0.229675(78)$ \\ 
$  4.0$ & $  0.013677(17)$ & $ -0.122890(42)$ & $ -0.109982(40)$ & $ -0.293550(90)$ \\ 
$  5.0$ & $  0.013371(18)$ & $ -0.154177(45)$ & $ -0.140336(48)$ & $ -0.35409(12)\,\ $ \\ 
$ 10.0$ & $  0.012716(15)$ & $ -0.304675(91)$ & $ -0.288888(86)$ & $ -0.63276(22)\,\ $ \\ 
$\infty$& $  0.012316(07)$ & $ -\infty      $ & $ -\infty      $ & $ -\infty       $ \\
\hline\hline
\end{tabular}
\caption{Numerical values of $Z_{M_2}^{[1]}$ 
for the SW action and the sD34 actions.}
\label{tab:zm2}
\end{center}
\end{table}

\clearpage 


\end{document}